\documentclass[lettersize,journal]{IEEEtran}
\usepackage{amsmath,amsfonts, amsthm, mathrsfs}
\usepackage{dsfont}
\usepackage{matlab-prettifier}
\usepackage{algpseudocode}
\usepackage{algorithm}
\usepackage{array}
\usepackage[caption=false,font=normalsize,labelfont=sf,textfont=sf]{subfig}
\usepackage{textcomp}
\usepackage{stfloats}
\usepackage{url}
\usepackage{verbatim}
\usepackage{graphicx}
\usepackage{cite}
\usepackage{xcolor}
\usepackage{arydshln}
\usepackage{cleveref}
\newcommand{\mc}[1]{\mathcal{#1}}
\newcommand{\mr}[1]{\mathrm{#1}}
\newcommand{\ms}[1]{\mathscr{#1}}

\newcommand{\mb}[1]{\mathbb{#1}}

\newcommand{\Matrix}[3][1.25]{\ensuremath{
		{\arraycolsep= #1 pt \def\arraystretch{1}
			\left[ \begin{array}{#2} #3 \end{array} \right]}}}

\newtheorem{definition}{Definition}
\newtheorem{theorem}{Theorem}
\newtheorem{lemma}{Lemma}

\newtheorem{example}{Example}

\begin{document}

\title{Iteratively Convex Frequency Data-Driven Synthesis of Structured MIMO Controllers}%: Applied to an Industrial Wirebonder}

\author{L. M. Spin, T. Samuels, D. Kosti\'{c}, W. P. M. H. Heemels, N. van de Wouw, R. T\'oth% <-this % stops a space
 \thanks{ %XXXX (project name) is funded by Holland High Tech | TKI HSTM via the PPP Innovation Scheme (PPP-I) for public-private partnerships.\\
	This work is co-financed by Holland High Tech, top sector High-Tech Systems and Materials, with a PPP innovation subsidy for public-private partnerships for research and development. %This work was also supported by the European Union within the framework of the National Laboratory for Autonomous 	Systems (RRF-2.3.1-21-2022-00002).
	Maurice Heemels has also received funding from the
	European Research Council under the Advanced ERC Grant Agreement PROACTHIS, no. 101055384.
} 
\thanks{L. M. Spin and R. T\'oth are with the Control Systems Group, N. van de Wouw is with the Dynamics and Control Group, W. P. M. H. Heemels is with the Control Systems Technology Group, Eindhoven University of Technology, The Netherlands. D. Kosti\'{c} and T. Samuels are with ASMPT Center of Competence, Beuningen, The Netherlands. R. T\'oth is also with the Institute for Computer Science and Control, Budapest, Hungary. Corresponding author: L. M. Spin, email address: \texttt{l.m.spin@tue.nl}.}}

% The paper headers
%\markboth{Journal of \LaTeX\ Class Files,~Vol.~14, No.~8, August~2021}%
%{Shell \MakeLowercase{\textit{et al.}}: A Sample Article Using IEEEtran.cls for IEEE Journals}

%\IEEEpubid{0000--0000/00\$00.00~\copyright~2021 IEEE}
% Remember, if you use this you must call \IEEEpubidadjcol in the second
% column for its text to clear the IEEEpubid mark.

\maketitle

\begin{abstract}
	Frequency-response-function-based controller design enables synthesis directly from measured data, avoiding the need for accurate parametric plant models. This paper presents a generalized IQC-based approach for convex synthesis of structured MIMO feedback and feedforward controllers from frequency response function data. The framework supports flexible controller parametrizations, multiple performance objectives, and both singular-value and element-wise performance shaping of MIMO controllers within a unified frequency-domain formulation. To facilitate practical deployment, the synthesis approach is embedded within a bootstrapped controller design methodology that progressively increases controller flexibility while refining the performance specifications. The resulting approach is validated experimentally on the motion stage of an industrial wirebonder, where clear performance improvement is demonstrated through MIMO $\mathcal{H}_\infty$ synthesis, element-wise feedback shaping, and dynamic feedforward design. The results demonstrate that high-performance multivariable controllers can be synthesized directly from measured FRF data in a practical and systematic manner.
\end{abstract}


\section{Introduction}\label{sec:intro}

Frequency-response-function (FRF)-based controller design plays a central role in the high-precision mechatronics industry~\cite{oomen2018advanced}, where performance requirements are stringent and plant dynamics are often difficult to capture accurately in parametric form~\cite{evers2020frequency, yang2012linear, van2018frequency, pintelon2012system}. In applications, such as semiconductor manufacturing equipment, precision positioning stages, and advanced motion systems, the dynamics are characterized by flexible modes, cross-axis dynamic coupling, and operating-point dependent dynamics~\cite{butler2011position}. In such settings, frequency-domain measurements provide a direct and reliable description of the relevant input--output behavior, while physics-based modeling or state-space identification can be time-consuming and fragile (i.e., prone to modelling errors). As a result, FRF-based control design has become an indispensable tool in industrial practice~\cite{heertjes2025stage, heertjes2020control, caporale2024practical}.

High-precision motion control systems (HPMCS) form a particularly demanding system class within this domain. These systems must achieve extreme positioning accuracy and high throughput simultaneously~\cite{steinbuch2013design}. From a control perspective, this translates into three key design objectives: (i) fast and reliable controller synthesis to enable on-site tuning and iterative refinement, (ii) high closed-loop bandwidth while respecting robustness constraints such as modulus margin requirements~\cite{lee1996robust}, and (iii) active dynamic decoupling of multivariable interactions to ensure uniform performance across all motion axes~\cite{falb1967decoupling, weekers2025unified, weekers2024data}. These requirements place stringent demands on both the controller structure and the underlying synthesis methodology.

Industrial wirebonders, as considered in this paper, serve as a representative benchmark for this class of systems. Such systems combine stringent performance requirements, significant multivariable interactions, and complex flexible dynamics, thereby providing a realistic application to assess advanced multivariable control design methodologies.

For multivariable systems, however, controller synthesis based on frequency-domain data remains challenging. Nonconvex optimization problems arising from FRF-based MIMO control design typically scale poorly with the number of optimization variables and system dimension and often lead to locally optimal solutions, whose performance and robustness are highly sensitive to initialization~\cite{apkarian2006nonsmooth, apkarian2018structured}. In practice, such solutions are rarely acceptable for high-precision applications, where repeatability, reliability, and predictable performance margins are essential. This has motivated a strong interest in convex synthesis approaches~\cite{scherer2000design, scherer2002structured}, which offer scalability, reproducibility, and the ability to systematically incorporate structural and performance constraints.

Within this context, Integral Quadratic Constraints (IQCs) have emerged as a particularly attractive framework for frequency-domain analysis and synthesis~\cite{megretski1997system, scherer2000linear}. IQCs provide a unified language to describe uncertainty, performance objectives, and system interconnections directly in the frequency domain~\cite{veenman2016robust}. Moreover, the associated frequency-domain inequalities (FDIs) naturally admit convex formulations, making IQCs especially well suited for controller synthesis directly from FRF data~\cite{scherer2022dissipativity}. %In this paper, we build on this paradigm and generalize it in several directions motivated by practical controller design requirements in HPMCS.

Despite these developments, several limitations remain in existing FRF-based and IQC-based synthesis methods. In addition, from a controller-design perspective, many approaches rely on fixed or low-complexity controller structures and provide limited support for systematically exploring increasingly sophisticated controller architectures. From a performance-design perspective, specifications are typically formulated in terms of singular-value bounds, providing limited control over individual input--output channels and dynamic decoupling objectives. Furthermore, weighting filters are often restricted to realizable LTI systems~\cite{zhou1996robust, boeren2013enhancing}, preventing the use of practically relevant non-smooth magnitude specifications that frequently arise in industrial applications.

The first contribution of this paper concerns controller parametrization. We introduce a flexible LTI controller structure that allows the designer to impose centralized or decentralized architectures by directly constraining individual elements of the MIMO controller transfer function. The controller order can be selected independently of the frequency grid, and practical features such as integral action and high-frequency roll-off can be incorporated explicitly. Beyond feedback control, the parametrization naturally extends to feedforward design and feedback--feedforward codesign.

The second contribution concerns performance specification and shaping. In addition to standard performance measures~\cite{antoulas2005approximation, scherer2006robust}, the proposed approach incorporates $\mc{H}_{\infty}$, $\mc{H}_2$, and $\mc{H}_1$ objectives through suitable performance multipliers and associated IQCs~\cite{nicoletti2019data, schuchert2024data}. Furthermore, as a significant advantage, element-wise performance specifications can be imposed directly on MIMO transfer functions, enabling channel-specific shaping and dynamic decoupling objectives~\cite{van2002multivariable, boeren2015non}. To facilitate efficient performance shaping, weighting function descriptions are effectively incorporated into the performance multipliers and the cost function, instead of the standard approach of including them in the plant, which allows the use of non-LTI weighting descriptions such as discontinuous or piecewise-defined magnitude bounds in the frequency domain.

The third contribution concerns the controller design methodology. Rather than focusing solely on synthesis conditions, the proposed framework is embedded within a practical bootstrapped design workflow tailored to HPMCS. Starting from diagonal SISO controllers reflecting standard industrial practice, controller complexity and performance requirements are progressively refined through MIMO $\mc{H}_{\infty}$ synthesis and element-wise performance shaping. The framework is further extended to feedforward design. 

Finally, the effectiveness of the developed approach is demonstrated experimentally on an industrial wirebonder. This validation culminates in a practical frequency-data-driven controller design methodology that combines flexible controller parametrizations, advanced performance shaping capabilities, and iteratively convex IQC-based synthesis within a unified approach.

The remainder of this paper is organized as follows. Section~\ref{sec:prob} formulates the controller design problem. Sections III--V introduce the IQC framework, performance formulations, and controller parametrizations for the proposed approach, respectively. Section~\ref{sec:syn} combines these ingredients into an iteratively convex synthesis procedure. Section~\ref{sec:design} embeds the synthesis method within a practical controller design workflow and demonstrates its applicability through an industrial wirebonder case study. Finally, Section~VIII presents the conclusions.

\noindent
{\bf Notation:}
The sets $\mb{C}$, $\mb{R}$, $\mb{Z}$, and $\mb{N}$ denote the complex numbers, real numbers, integers, and non-negative integers, respectively. Matrices of dimension $n\times m$ over a field $\mb{K}$ are denoted by $\mb{K}^{n\times m}$, while $\mb{K}^{\bullet\times\bullet}$ indicates arbitrary dimensions. The set of real symmetric and complex Hermitian matrices are denoted by $\mb{S}^{\bullet}$ and $\mb{H}^{\bullet}$, respectively. We write $A\succ0$ ($A\succeq0$) and $A\prec0$ ($A\preceq0$) for positive (semi-)definite and negative (semi-)definite Hermitian matrices, respectively, and denote the set of positive definite real symmetric matrices by $\mb{S}_{\succ0}^{\bullet}$. The open unit disc, unit circle, closed unit disc, and exterior of the closed unit disc are denoted by $\mb{D}$, $\partial\mb{D}$, $\bar{\mb{D}}$, and $\mb{E}$, respectively. Furthermore, let $\Omega:=[-\pi,\pi]$. For a discrete-time LTI system $G$, the associated transfer function is denoted by $G(z)$. We write $G\in\mc{R}_{\mr p}^{\bullet\times\bullet}$, if $G$ is proper and real-rational, and $G\in\mc{RH}_{\infty}^{\bullet\times\bullet}(\mb{E})$ if it is analytic on $\mb{E}$ and bounded on $\partial\mb{D}$. Transfer matrices with bounded spectra on $\partial\mb{D}$ are denoted by $\mc{RL}_{\infty}^{\bullet\times\bullet}(\partial\mb{D})$, while $\mc{L}_{\infty}^{\bullet\times\bullet}(\partial\mb{D})$ denotes the set of essentially bounded matrix-valued (complex) functions on $\partial\mb{D}$. For two transfer matrices $G_1$ and $G_2$, $G_1\star G_2$ denotes the Redheffer product \cite{zhou1996robust}. Finally, if $G(e^{j\omega})\prec0$ for all $\omega\in\Omega$, we write $G\prec0$ on $\partial\mb{D}$. %Finally, $\ell_2^{\bullet\times\bullet}(\mb{N})$ and $\ell_{2\mr e}^{\bullet\times\bullet}(\mb{N})$ denote the spaces of square-integrable and extended square-integrable discrete-time signals, respectively.

\section{Problem Formulation}\label{sec:prob}
\begin{figure}[b]
	\vspace*{-15pt}
	\centering
	\includegraphics[width=0.9\linewidth]{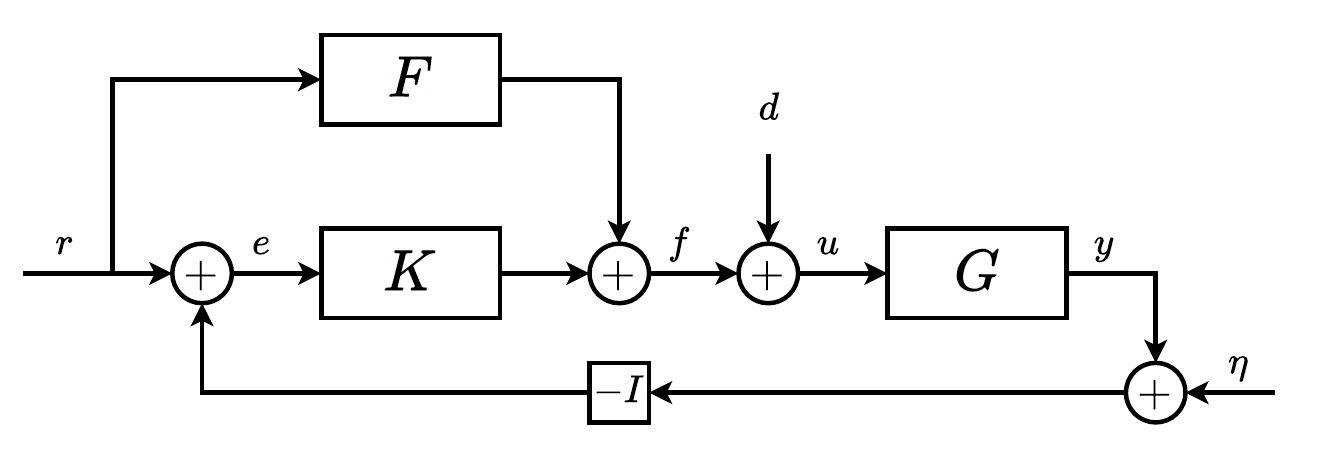}
	\caption{Feedback and Feedforward control configuration.}
	\label{fig:FFandFBControlConfig}
\end{figure}
Consider the discrete-time LTI plant, described by
$G\in\mc{RL}_{\infty}^{n_y\times n_u}$, for which information is assumed to be
available only through frequency-response measurements $G(e^{j\omega})$. In this paper, we consider the feedback configuration shown in
Fig.~\ref{fig:FFandFBControlConfig} in which $K$ denotes a feedback controller and $F$ a feedforward controller. The control configuration is defined by the following signal interconnections:
\begin{subequations}
	\begin{align}
		u &= f+d, &
		f &= Ke + Fr,
		\\
		e &= r-\eta-y, &
		y &= Gu,
	\end{align}
	\label{eq:closedloop}
\end{subequations}
where $r$ denotes the reference signal, $d$ an input
disturbance, $\eta$ measurement noise, $e$ the tracking
error, $f$ the controller output, $u$ the plant input, and
$y$ the plant output.

For the feedback-only configuration
$(F=0)$, the closed-loop mapping from $[r^\top\; -d^\top]^\top$ to $[e^\top\; f^\top]^\top$ is characterized
by the output sensitivity functions
\begin{equation}
	\Matrix{cc}{S & SG \\ KS & GSK}, \quad 
	S = (I+GK)^{-1},
	\label{eq:ST}
\end{equation}
whereas, when feedforward action is included, the
reference-to-error transfer function is given by $S(I-GF)$. These transfer functions capture disturbance rejection, robustness, and reference-tracking
performance, and therefore form the primary
closed-loop functions considered throughout this paper.

The objective is to synthesize feedback and feedforward
controllers directly from measured FRF data such that the
resulting closed-loop functions satisfy prescribed performance
requirements. Throughout this paper, these requirements are
expressed in terms of weighted versions of the closed-loop transfer
functions, allowing disturbance
rejection, robustness, tracking performance, and dynamic
decoupling objectives to be addressed within a common
frequency-domain approach.

Since controller synthesis for multivariable systems is
generally non-convex and scales poorly with controller
complexity, the focus of this paper is on (iteratively) convex synthesis
formulations. Consequently, both the closed-loop stability
conditions and the performance specifications must admit
convex reformulations with respect to the controller
parameters. To facilitate such reformulations, the feedback controller is
represented through a right coprime factorization
\begin{equation}
	K = UV^{-1},
	\label{eq:coprime}
\end{equation}
where the stable factors $U$ and $V$ belong to a prescribed
controller class. This parametrization naturally accommodates
a wide variety of controller structures, as detailed in Section~\ref{sec:controller}.%centralized and decentralized architectures, integral action,
%high-frequency roll-off, and user-defined controller
%structures while preserving the algebraic structure required
%for convex synthesis. 
The same construction can be applied to the feedforward controller whenever feedforward design is considered.

The remainder of this paper develops an IQC-based
frequency-domain approach that enables convex analysis and
synthesis directly from measured FRF data. First, IQC-based
stability and performance conditions are introduced. Next,
performance objectives and controller parametrizations are
formulated in a manner suitable for convex optimization.
Finally, these ingredients are combined into a practical
bootstrapped controller design methodology and validated on
an industrial wirebonder.

\section{IQC-Based Frequency-Domain approach}\label{sec:prelim}

The proposed synthesis method is formulated within the
Integral Quadratic Constraint (IQC) framework
\cite{megretski1997system,veenman2016robust}. IQCs provide a
unified frequency-domain description of uncertainty and
performance through quadratic constraints on signal pairs. Here we provide the definition of IQCs for signals.

\begin{definition}[Integral Quadratic Constraint]
	For frequency-domain signals $(q,p) \in \mathscr{L}_2^{n_\mr{q}+n_\mr{p}}(\partial \mathbb{D})$ and a multiplier $\Pi = \Pi^* \in \ms{L}_\infty^{n_\mr{q}+n_\mr{p} \times n_\mr{q}+n_\mr{p}} (\partial \mb{D})$ with partitioning
	\begin{equation*}
		\Pi := \Matrix{cc}{\Pi_\mr{qq} & \Pi_\mr{qp} \\ \Pi_\mr{pq} & \Pi_\mr{pp}},
	\end{equation*}
	define the quadratic functional 
	\begin{equation}
		\mc{I}(\Pi, q, p):=\int_{-\pi}^{\pi}
		\begin{bmatrix}
			q(e^{j\omega})\\
			p(e^{j\omega})
		\end{bmatrix}^*
		\Pi(e^{j\omega})
		\begin{bmatrix}
			q(e^{j\omega})\\
			p(e^{j\omega})
		\end{bmatrix}
		\, d\omega.
	\end{equation}
	The pair $(q,p)$ is said to satisfy an IQC defined by $\Pi$, if $\mc{I}(\Pi,q, p) \geq 0$.
\end{definition}

We now use IQCs to analyze interconnected systems whose
internal signals are constrained by a common uncertainty
description. Consider the uncertain interconnection
$T_{zw}=\Delta\star M$, where
$\Delta\in\boldsymbol{\Delta}$ is a causal bounded operator
belonging to an uncertainty class $\boldsymbol{\Delta}$ and
$M\in\mc{RH}_{\infty}$ is a stable LTI system. The system
$M$ maps the uncertainty output $p$ and exogenous input $w$
to the uncertainty input $q$ and performance output $z$,
i.e.,
\begin{equation}
	\Matrix{c}{q \\ z}
	=
	M
	\Matrix{c}{p \\ w}
	=
	\Matrix{cc}{M_{\mr{qp}} & M_{\mr{qw}} \\ M_{\mr{zp}} & M_{\mr{zw}}}
	\Matrix{c}{p \\ w}.
	\label{eq:UncertainInterconnection}
\end{equation}
The resulting closed-loop map from $w$ to $z$ is denoted by
$T_{zw}=\Delta\star M$.

Rather than analyzing every $\Delta\in\boldsymbol{\Delta}$
separately, the uncertainty class is characterized by a
multiplier set $\boldsymbol{\Pi}$. More precisely, each
$\Delta\in\boldsymbol{\Delta}$ is assumed to satisfy the IQC
defined by every $\Pi\in\boldsymbol{\Pi}$, i.e., $\mc{I}(\Pi,q,\Delta(q))\geq0$. In this paper, the multiplier set is chosen as a subset of the
inertia constraints multiplier class~\cite{veenman2013stability, seiler2010dissipation}
\begin{multline}\label{eq:PNMultiplierClass}
	\boldsymbol{\Pi}_\mr{in} := \Bigg\{ \Pi = \Pi^* = \Matrix{cc}{\Pi_\mr{qq} & \Pi_\mr{qp} \\ (\Pi_\mr{qp})^* & \Pi_\mr{pp}} \in \mc{L}_\infty^{\bullet \times\bullet} (\partial \mb{D}) \ \Bigg| \\ \Pi_\mr{qq} \succ 0 \ \mr{on} \ \partial \mb{D}, \ \Pi_{pp} \preceq 0 \ \mr{on} \ \partial \mb{D} \Bigg\}.
\end{multline}
The performance specification on the closed-loop
$T_{zw}$ is encoded through a second multiplier
$\Pi_{\mr{p}}$, acting on the performance channel $(z,w)$, whose construction is discussed in Section~\ref{sec:perf}. The following frequency-domain condition is used throughout the paper to certify robust stability and performance~\cite{veenman2016robust}.

\begin{theorem}[Robust Performance]\label{thm:IQC}
	Suppose that the interconnection $\Delta\star M$ is well
	posed for all $\Delta\in\boldsymbol{\Delta}$ and that all
	$\Delta\in\boldsymbol{\Delta}$ satisfy
	$\mc{I}(\Pi,q,\Delta(q))\geq0$ for all
	$\Pi\in\boldsymbol{\Pi}$. If there exist
	$\Pi\in\boldsymbol{\Pi}$ and a performance multiplier
	$\Pi_{\mr{p}}$ such that
	\begin{equation}
		\Matrix{cc}{M_{\mr{qp}} & M_{\mr{qw}} \\ I & 0 \\ M_{\mr{zp}} & M_{\mr{zw}} \\ 0 & I}^{*}
		\Matrix{cc}{\Pi & 0 \\ 0 & \Pi_{\mr{p}}}
		\Matrix{cc}{M_{\mr{qp}} & M_{\mr{qw}} \\ I & 0 \\ M_{\mr{zp}} & M_{\mr{zw}} \\ 0 & I}
		\prec 0,
		\label{eq:RobustPerformanceFDI}
	\end{equation}
	for all $\omega\in\Omega$, then
	$T_{zw}=\Delta\star M$ is robustly stable and satisfies the
	performance specification encoded by $\Pi_{\mr{p}}$ for all
	$\Delta\in\boldsymbol{\Delta}$.
\end{theorem}

The inequality \eqref{eq:RobustPerformanceFDI} is referred to
as a frequency-domain inequality (FDI). In the sequel, the performance multiplier $\Pi_{\mr{p}}$ and its associated cost function $J(\Pi_{\mr{p}})$ are introduced
to encode the desired closed-loop performance objectives.
Different choices of $\Pi_{\mr{p}}$ and $J(\Pi_{\mr{p}})$
give rise to different performance specifications and
optimization criteria.

\section{Performance Characterization}\label{sec:perf}

In this section, we discuss several performance metrics in the frequency domain which can be optimized for by choosing different performance multiplier classes $\boldsymbol{\Pi}^\mr{p}$ and associated cost functions. Towards this end, we introduce the Hardy spaces~\cite{antoulas2005approximation}
\begin{equation}\label{eq:HpSpace}
	\!\!\!\mathcal{H}_p^{n \times m}(\mb{E}) :=
	\left\{
	X : \mathbb{E} \rightarrow \mathbb{C}^{n \times m}
	\;\middle|\!\!
	\begin{array}{l}
		X \text{ is analytic in } \mathbb{E}, \\
		\|X\|_{\mathcal{H}_p} < \infty
	\end{array}
	\!\!\!\right\},
\end{equation}
where the frequency-domain norm $\|\cdot\|_{\mc{H}_p}$ is defined as
\begin{equation}
	\!\!\!\!\!\!\|X\|_{\mc{H}_p} = \left\{ \begin{aligned}
		\bigg( \frac{1}{2\pi} \int_\Omega &\|X(e^{j \omega})\|^p_{S_p} \ \mr{d} \omega \bigg)^{\frac{1}{p}}, \!\!&&p \in [1, \infty),\\
		\sup_{\omega\in\Omega}&\|X(e^{j \omega})\|_{S_\infty}, &&p = \infty.
	\end{aligned}\right. \label{eq:HpNorm}
\end{equation}
Here $\| A \|_{S_p}$ denotes the Schatten $p$-norm of a matrix $A$, e.g., the vector $p$-norm of the singular values $\sigma_i$ of $A$:
\begin{equation}
	\|A \|_{S_p} = \left( \sum_{i = 1}^{\mr{rank}(A)} \sigma_i^p(A) \right)^\frac{1}{p}, \ \|A\|_{S_\infty} = \sigma_1(A).
\end{equation}
For rational LTI systems, the Hardy space $\mathcal{H}_p$ corresponds to transfer functions with all poles strictly outside the open unit disk, such that their frequency response $X(e^{j\omega})$ is well-defined on $\partial \mathbb{D}$.
The frequency-domain norms of the FRF of $X$ can also be interpreted in the time-domain, see~\cite{nicoletti2019data, scherer2000linear}. However, we can directly evaluate performance on the FRF $G(e^{j \omega})$ by evaluating~\eqref{eq:HpNorm} if this can be cast as a semi-definite program (SDP) in the presented IQC framework for controller synthesis. This can be done, provided the matrix norms can be written as an SDP and as an IQC performance multiplier $\Pi^\mr{p}$ simultaneously. In the following subsection, we will discuss how this can be achieved for different matrix norms $\|\cdot \|_{S_p}$.
\subsection{IQC-based performance formulation}
Consider a closed-loop system $T_\mr{zw}$ which maps exogenous input $w$ to performance output $z$. We aim at deriving convex formulations of performance with respect to the performance multiplier $\Pi^\mr{p}$. Toward that end, performance is given in terms of the IQC on the performance channel $(z, w)$
\begin{equation}\label{eq:perfIQC}
	\mc{I}(\Pi^\mr{p}, z, w) \leq -\varepsilon \| w\|_{\ms{L}_2}^2, \ \varepsilon > 0.
\end{equation}
In particular, we will work with the following performance multiplier class parametrization
\begin{equation}
	\boldsymbol{\Pi}_\mr{p} := \left\{ \Pi^\mr{p} \in \boldsymbol{\Pi}_\mr{in} \ \bigg | \ \Pi^\mr{p} = \Matrix{cc}{Q_\mr{p}^{-1} & 0 \\ 0 & -R_\mr{p}} \right\},
\end{equation}
leading to an FDI formulation of~\eqref{eq:perfIQC}:
\begin{equation}\label{eq:perfFDI}
	T_\mr{zw}^* Q_\mr{p}^{-1} T_\mr{zw}-R_\mr{p} \prec 0, \ \mr{on} \ \partial \mb{D}.
\end{equation}
Structural constraints of $Q_\mr{p}$ and $R_\mr{p}$, together with a suitable cost function $J(\Pi^\mr{p})$, provide Hardy-norm performance objectives. These performance objectives are summarized as, see~\cite{scherer2000linear}:
\begin{itemize}
	\item \textbf{$\mc{H}_\infty$:} 
	$Q_p = \gamma I_{n_z}$, $R_p = \gamma I_{n_w}$, with $J_{\mc{H}_\infty} = \gamma$;
	
	\item \textbf{$\mc{H}_2$:} 
	$Q_p = I_{n_z}$, $R_p \succ 0$ on $\partial \mb{D}$, with
	\[
	J_{\mc{H}_2}^2 = \frac{1}{2\pi} \int_{-\pi}^{\pi} \mathrm{Tr}\big(R_p(e^{j\omega})\big)\, d\omega;
	\]
	
	\item \textbf{$\mc{H}_1$:} 
	$Q_p \succ 0$, $R_p \succ 0$ on $\partial \mb{D}$, with
	\[
	J_{\mc{H}_1} = \frac{1}{2\pi} \int_{-\pi}^{\pi} 
	\frac{\mathrm{Tr}\big(Q_p(e^{j\omega})\big) + \mathrm{Tr}\big(R_p(e^{j\omega})\big)}{2}
	\, d\omega.
	\]
\end{itemize}

The cost functions are linear functionals of the optimization variables $(Q_\mr{p}, R_\mr{p})$, however the FDI~\eqref{eq:perfFDI} contains an inverse which prohibits SDP based optimization of this problem. Applying Schur's complement pointwise over the spectrum would convexify this problem. However, as will become apparent in Section~\ref{sec:syn}, an alternative convexification is required as well, which is provided in the following lemma.

\begin{lemma}[Convex FDI]\label{lemma:Slack}
	Let $T_\mr{zw},Q_\mr{p},R_\mr{p} \in \mathscr{L}_\infty(\partial\mathbb{D})$ be compatible Hermitian matrix-valued functions with $Q_\mr{p} \succ 0$ and $R_\mr{p} \succ 0$ on $\partial\mathbb{D}$. Then, the FDI~\eqref{eq:perfFDI}
	%\begin{equation}\label{eq:nonConvexPerfMultFDI}
	%	\Matrix{c}{T_{zw} \\ I}^*
	%	\Matrix{cc}{Q^{-1} & 0 \\ 0 & -R}
	%	\Matrix{c}{T_{zw} \\ I}
	%	\prec 0
	%	\ \ \mr{on} \ \  \partial\mathbb{D}
	%\end{equation}
	holds if and only if there exists a Hermitian slack variable
	$S \in \mathscr{L}_\infty(\partial\mathbb{D})$, having the same structure as $Q_\mr{p}$, such that
	\begin{equation}\label{eq:normUniversalPerfMultFDI}
		\Matrix{c}{T_{zw} \\ I}^*
		\Matrix{cc}{S & 0 \\ 0 & -R_\mr{p}}
		\Matrix{c}{T_{zw} \\ I}
		\prec 0,
		\ \
		\Matrix{cc}{Q_\mr{p} & I \\ I & S}
		\succeq 0
		\ \ \mr{on}  \ \ \partial\mathbb{D}.
	\end{equation}
	Hence, the feasibility of \eqref{eq:perfFDI} can be verified through the convex constraints in \eqref{eq:normUniversalPerfMultFDI}.
\end{lemma}
\begin{proof}
	The sufficiency follows by choosing $S=Q_\mr{p}^{-1}$, for which
	\[
	\Matrix{cc}{Q_\mr{p} & I \\ I & Q_\mr{p}^{-1}} \succeq 0  \ \mr{on}  \ \partial\mathbb{D},
	\]
	holds by Schur's complement. Conversely,
	\[
	\Matrix{cc}{Q_\mr{p} & I \\ I & S}
	\succeq 0  \ \mr{on}  \ \partial\mathbb{D}
	\iff
	S-Q_\mr{p}^{-1}\succeq0  \ \mr{on}  \ \partial\mathbb{D},
	\]
	and therefore
	\[
	T_\mr{zw}^*Q_\mr{p}^{-1}T_\mr{zw}-R_\mr{p}
	\preceq
	T_\mr{zw}^*ST_\mr{zw}-R_\mr{p}
	\prec0 \ \mr{on}  \ \partial\mathbb{D},
	\]
	proving the claim.
\end{proof}

By introducing positive defininite slack variables, the performance analysis of $\mathcal{H}_p$-norms can be cast as SDPs within the IQC framework. Now that the IQC-based machinery for Hardy norm analysis is defined, we will discuss the weighted performance problem, for shaping purposes, in the next subsection.

\subsection{Weighting filters and performance shaping}
So far, we have not considered weighting filters on the inputs and outputs of the system $T_{zw}$ in the forms of $W_w$ and $W_z$, respectively, see \cite{van2002multivariable}. These weighting filters provide a means to shape the desired closed-loop performance across frequency by emphasizing or relaxing the performance requirements in selected frequency ranges, for which established design guidelines are available, see, e.g., \cite{skogestad2007multivariable}. Therefore, we would like to use the weighted performance objective formulated as
\begin{equation}
	\|W_z T_\mr{zw} W_w\|_{\mc{H}_p}.
\end{equation}
However, IQC-based performance analysis, as discussed in Theorem~\ref{thm:IQC}, works provided that $W_z$, $T_{zw}$, and $W_w$ are elements of $\mc{RH}_\infty^{\bullet \times \bullet}$. This, however, is limiting in practice. Standard weighting-filter designs typically employ LTI filters and therefore satisfy this requirement. While such filters enable conventional performance shaping, maximizing performance can benefit from directly prescribing frequency-dependent magnitude bounds that need not correspond to an LTI system and may, in particular, be discontinuous. Such discontinuous bounds cannot be expressed as LTI systems. Hence, such bounds cannot be incorporated into the FDI~\eqref{eq:perfFDI} directly. In order to circumvent this issue, we propose to solve an equivalent problem by means of a change of variables that effectively relocates the (discontinuous) weighting filters to the cost function thereby creating a weighted SDP variant with new optimization variables. Let us define $\widetilde{Q}_p$ and $\widetilde{R}_p$ which have the same structural properties, e.g. block-diagonal structure, as $Q_p$ and $R_p$, respectively. These new variables satisfy the relation
\begin{equation}\label{eq:perfVarChange}
	\widetilde{R}_p = (W_w^*)^{-1} R_p W_w^{-1}, \quad \widetilde{Q}_p = W_z^{-1} Q_p (W_z^*)^{-1},
\end{equation}
for which it holds that
\begin{equation}
	\Matrix{c}{W_z T_{zw} W_w \\ I}^* \Pi_p \Matrix{c}{W_z T_{zw} W_w \\ I} \prec 0 \ \ \mr{on}  \ \ \partial\mathbb{D},
\end{equation}
is  feasible if and only if
\begin{equation}
	\Matrix{c}{T_{zw}\\ I}^* \widetilde{\Pi}_p \Matrix{c}{T_{zw} \\ I} \prec 0 \ \ \mr{on}  \ \ \partial\mathbb{D},
\end{equation}
is feasible, where $\widetilde{\Pi}_p = \mr{diag}(\widetilde{Q}^{-1}_p, \widetilde{R}_p)$. However, the cost function is still a function of the original variables $Q_p$ and $R_p$; hence we apply~\eqref{eq:perfVarChange} and note that we can make the cost function convex in the new variables as
\begin{equation}
	J(Q_p, R_p) = J(W_z \widetilde{Q}_p W_z^*, \ W_w^* \widetilde{R}_p W_w).
\end{equation}
The IQC multiplier $\widetilde{\Pi}_p$ is only required to be a measurable, essentially bounded, Hermitation-valued function instead of a realization of an LTI system which is why this change of variables is allowed within the IQC framework. For $\mathcal{H}_\infty$, we have diagonal structure of the original variables, but in order to have the same structure in the new variables while simulatenously respecting the relation~\eqref{eq:perfVarChange} the weighting filters also are required to be diagonal. This restriction does not apply to $\mathcal{H}_2$ and $\mathcal{H}_1$.

The Hardy norms are functionals of the frequency-dependent singular values of the FRF. The shaping of singular values is highly non-intuitive and engineers prefer to shape individual elements of an FRF instead of shaping singular values. This objective is pursued in the following subsection.

\subsection{Element-wise shaping}
In order to focus on elements of $T_{zw}$ instead of its singular values, we start by defining $T_{ij} = e_i^\top T_{zw} e_j$ for $i = 1, \cdots, n_z,$ and $j = 1, \cdots, n_w,$ and where $e_i$ denotes the basis vector with a 1 on the $i$-th place and zero elsewhere. Individual weighting filters can then be imposed by replacing the single FDI~\eqref{eq:perfFDI} with the collection of FDIs%\footnote{Each additional constraint, related to each element $(T_{zw})_{i,j}$, does have a reduced dimension, but the overall dimension of all constraint increases significantly likely resulting in numerical trouble.}
\begin{equation}\label{eq:elementPerf}
	\Matrix{c}{T_{ij}\\ I}^* \widetilde{\Pi}_{p, ij} \Matrix{c}{T_{ij} \\ I} \prec 0 \ \ \mr{on}  \ \ \partial\mathbb{D},
\end{equation}
for $i = 1, \cdots, n_z,$ and $j = 1, \cdots, n_w.$ Associated with each element-wise performance constraint, define the corresponding cost function $J_{ij} = J(\widetilde{\Pi}_{p, ij})$ and impose the upper bounds $J_{ij} \leq \gamma_{ij}$. Collecting the variables $\gamma_{ij}$ into the matrix
\begin{equation}
	\Upsilon = (\gamma_{ij})_{i=1,\cdots, n_\mr{z}}^{j=1,\cdots, n_\mr{w}} \in \mb{R}^{n_\mr{z} \times n_\mr{w}},
\end{equation}
allows the overall optimization objective to be expressed, for example, as
\begin{equation}
	J = \max_{ij} \gamma_{ij} \quad \mr{or} \quad J = \|\Upsilon\|_{S_2}.
\end{equation}

To summarize, we have generalized $\mathcal{H}_p$ performance in the frequency domain using IQCs and have enabled the use of discontinuous weighting function as well as element-wise shaping of FRFs in contrast to singular value shaping, where such element-wise shaping provides much more flexibility in practice. Moreover, if a state-space representation of the model is available, then the FDI can be reformulated, through the KYP Lemma, as a single LMI~\cite{veenman2016robust}. However, there are no known analogues results for $\mathcal{H}_1$ performance, non-LTI weighting filters, and mixed element-wise cost function which emphasizes the flexibility that this frequency domain formulation offers compared to state-space methods.

In the next section, we introduce coprime-based parametrizations of the controller to be tuned. This is the last ingredient of our approach that we need to introduce before we can present the proposed controller synthesis algorithm.

\section{Controller parametrization}\label{sec:controller}
In this section, we will discuss the controller parametrization used in the controller synthesis approach outlined in Section~\ref{sec:prob}. Firstly, coprime factor-based controller parametrizations will be discussed for centralized MIMO design. Then, de-centralized controller structures will be discussed as well as enforcing integrators and high-frequency roll-off constraints will be discussed. Secondly, we provide conditions on coprimeness which are necessary for guaranteeing internal stability. This result will be used to add user-defined (not tunable) additional controller structure into (de)-centralized controller parametrization. Lastly, we will provide coprime factorization of most-used controller structures such as, PID, notches, low-pass filters, series and parallel interconnections.

\subsection{Coprime factorization}
Leading up to the iteratively convex synthesis algorithm, we parametrize the controller $K$ through stable coprime factors $UV^{-1}$. A pair $(U,V)$, with $U,V\in\mathcal{RH}_\infty(\mathbb E)$ and $V$ square, is called right coprime over $\mathcal{RH}_\infty(\mathbb E)$ if there exist $X,Y\in\mathcal{RH}_\infty(\mathbb E)$ such that the Bezout identity
\begin{equation}
	X V + Y U = I,
\end{equation}
holds, see~\cite{vidyasagar2022control}. Equivalently, the stacked factor $\begin{bmatrix} V^\top & U^\top \end{bmatrix}^\top$ admits a stable left inverse. A controller is then represented as $K = U V^{-1}$. The coprimeness condition excludes unstable pole-zero cancellations between $U$ and $V$, see~\cite{vidyasagar2022control}, and therefore ensures that unstable modes of the controller are represented explicitly through the zeros of $V$ rather than being hidden by cancellations.

This representation is particularly useful for black-box controller parametrizations, where only the controller order is prescribed. The factors $U$ and $V$ can be chosen to be stable and affine in the controller parameters, while the resulting controller $K=UV^{-1}$ may itself be unstable. Consequently, unstable stabilizing controllers can be represented without introducing unstable basis functions. The following subsection shows how this parametrization can be specialized to impose additional controller structure.

\subsection{Basis function parametrization of coprime factors}
We parametrize the stable factors $U$ and $V$ by generalized orthonormal basis functions over the Hardy space $\mathcal H_2(\mathbb E)$, following~\cite{van2000modelling,heuberger1995generalized,toth2010modeling}. Let
\begin{equation}
	\phi(z) = (zI-A_\phi)^{-1}B_\phi,
\end{equation}
denote a fixed vector of stable basis functions. The choice of $\phi$ is made by the user and may correspond, for example, to finite impulse-response, Laguerre, Kautz, Takenaka--Malmquist, or more general General Orthogonal Basis Functions (GOBFs) bases. The only property used in the sequel is that $\phi$ is stable and fixed, such that the coprime factors $U$ and $V$ can be parametrized as

%For a sufficiently rich basis, stable transfer matrices can be approximated as linear combinations of these basis functions. In particular, if the pole sequence underlying the Takenaka--Malmquist construction satisfies the usual completeness condition
%\begin{equation}
%	\sum_{k=1}^{\infty} (1-|p_k|)=\infty,
%\end{equation}
%then the corresponding orthonormal basis is complete in $\mathcal H_2(\mathbb E)$. Hence any stable square-integrable transfer function can be approximated arbitrarily well in the $\mathcal H_2$ norm by a finite expansion. In practice, we use a finite-dimensional truncation, which yields a tractable controller class whose approximation quality is determined by the selected poles and the number of basis functions.
\begin{equation}
	\begin{bmatrix}
		V \\ U
	\end{bmatrix}
	=
	\Matrix{cc}{D_V \\ D_U} + \Matrix{cc}{C_V \\ C_U} \phi,
\end{equation}
which are affine functions of the optimization variables $\theta := \mr{vec}(C_V,D_V,C_U,D_U)$. %Equivalently,
%\begin{equation}
%	V(z)=D_V+C_V\phi(z),\qquad
%	U(z)=D_U+C_U\phi(z).
%\end{equation}
Whenever $D_V$ is nonsingular, the resulting controller has the realization
\begin{equation}
	K=UV^{-1}
	=
	\left[
	\begin{array}{c|c}
		A_\phi-B_\phi D_V^{-1}C_V & B_\phi D_V^{-1} \\ \hline
		C_U-D_U D_V^{-1}C_V & D_U D_V^{-1}
	\end{array}
	\right].
\end{equation}
Structural constraints, such as a decentralized structure or sparsity-constrained controller architectures, can be imposed by enforcing corresponding linear constraints on the coefficient matrices $C_V,D_V,C_U,D_U$; see, e.g.,~\cite{schuchert2024data}.

\subsection{Controller structure}
Having introduced coprime factorizations and the GOBF parametrization of their factors, we now specialize this parametrization to structured controller architectures.
\subsubsection{(De)-Centralized structure}\noindent\\  \vspace*{-10pt}
\label{subsec:decen}

\noindent
For the right factorization of the controller, we use the following structure of the parameters $(V, U)$ to ensure (de)-centralized structure of the controller $K = UV^{-1}$. The GOBF parametrization can be chosen to obtain the following structure:
\begin{equation}
	A_\phi = \Matrix{ccc}{A_1 & 0 & 0 \\ 0 & A_2 & 0 \\ 0 & 0 & A_3}, \quad
	B_\phi = \Matrix{ccc}{B_1 & 0 & 0 \\ 0 & B_2 & 0 \\ 0 & 0 & B_3},
\end{equation}
which is a concatenation of three different GOBF bases. Therefore, we can impose a block-sparse structure on $K$ by using the following constraints on the free variables $C_V, C_U, D_V,$ and $D_U$. First, we also choose a block-diagonal structure on the denominator variables $C_V$ and $D_V$ as follows:
\begin{equation}
	C_V = \Matrix{ccc}{C_1 & 0 & 0 \\ 0 & C_2 & 0 \\ 0 & 0 & C_3}, \quad D_V = \Matrix{ccc}{D_1 & 0 & 0 \\ 0 & D_2 & 0 \\ 0 & 0 & D_3}.
\end{equation}
Secondly, we parametrize the numerator variables $C_U$ and $D_U$ as
\begin{equation}
	C_U = \Matrix{ccc}{C_{11} & C_{12} & C_{13} \\ C_{21} & C_{22} & C_{23} \\ C_{31} & C_{32} & C_{33}}, \quad D_U = \Matrix{ccc}{D_{11} & D_{12} & D_{13} \\ D_{21} & D_{22} & D_{23} \\ D_{31} & D_{32} & D_{33}}.
\end{equation}
These choices lead to the following structured coprime factors
%\begin{equation}
%	\begin{aligned}
%		C_V &= \Matrix{ccc}{C_1 & 0 & 0 \\ 0 & C_2 & 0 \\ 0 & 0 & C_3}\\
%		D_V &= \Matrix{ccc}{D_1 & 0 & 0 \\ 0 & D_2 & 0 \\ 0 & 0 & D_3}\\
%		C_U &= \Matrix{ccc}{C_{11} & C_{12} & C_{13} \\ C_{21} & U_{22} & C_{23} \\ C_{31} & C_{32} & C_{33}}\\
%		D_U &= \Matrix{ccc}{D_{11} & D_{12} & D_{13} \\ D_{21} & D_{22} & D_{23} \\ D_{31} & D_{32} & D_{33}}
%	\end{aligned}
%\end{equation}
\begin{equation}\label{eq:diagCon}
	V = \Matrix{ccc}{V_{1} & 0 & 0 \\ 0 & V_{2} & 0 \\ 0 & 0 & V_{3}}, \quad U = \Matrix{ccc}{U_{11} & U_{12} & U_{13} \\ U_{21} & U_{22} & U_{23} \\ U_{31} & U_{32} & U_{33}}.
\end{equation}
By imposing sparsity patterns on $C_U$ and $D_U$ directly induces the corresponding sparsity pattern in the coprime factor $U$. Combined with the block-diagonal structure of $V$, this yields the desired sparsity pattern of the controller $K = UV^{-1}$. %So, by appropriate choices of the coprime factors, a (de)-centralized structure for the controller $K$ can be obtained by sparsity constraints in the controller parameters.

\subsubsection{Integral and high-frequency roll-off control}\noindent\\  \vspace*{-10pt}

\noindent
Since the frequency response $X(e^{j \omega})$ of a real rational transfer function $X$ is real-valued for $\omega \in \{-\pi, 0, \pi\}$, these frequency points can be exploited to impose additional structural constraints on the controller parametrization. In particular, integral action is enforced in the (de)-centralized controller by imposing the constraint
%\begin{equation}
%	D_V = \begin{cases}
%		-C_V(I-A_\phi)^{-1} B_\phi, \ \mr{for \ a \ r.f. \ of} \ K, \\
%		-C_\varphi(I-A_\varphi)^{-1}B_V, \ \mr{for \ a \ l.f. \ of} \ K.
%	\end{cases}
%\end{equation}
\begin{equation}
	D_V = -C_V(I-A_\phi)^{-1} B_\phi,
\end{equation}
which yields $V(1) = 0$, forcing an integrator in $K(1)$. Similarly, high-frequency roll-off is enforced by imposing the constraint
%\begin{equation}
%	D_U = \begin{cases}
%		C_U(I+A_\varphi)^{-1} B_\varphi, \ \mr{for \ a \ r.f. \ of} \ K, \\
%		C_\varphi(I+A_\varphi)^{-1}B_U, \ \mr{for \ a \ l.f. \ of} \ K.
%	\end{cases}
%\end{equation}
\begin{equation}
	D_U = C_U(I+A_\phi)^{-1} B_\phi.
\end{equation}
which yields $U(-1) = 0$, forcing roll-off in $K(-1)$. Incorporating these constraints reduces the number of free controller parameters. 
%Depending entirely on $C_V, C_U, D_V,$ and $D_U$ the controller factors $V$ and $U$ might, or might not, be coprime. In the following subsection, we will discuss when a (right or left) factorization is also a coprime factorization and how this notion related to internal stability.
The following subsection specializes the proposed parametrization to second-order filters
%\subsubsection{Canonical filter structures}\noindent\\  \vspace*{-10pt}
\subsubsection{Canonical filter structures}\noindent\\[-10pt]

\noindent
Many controller structures encountered in industrial motion control, including PID, low-pass, lead, lag, and notch filters, can be represented as special cases of a canonical second-order infinite impulse response (IIR) filter. Rather than deriving a separate coprime parametrization for each controller architecture, we first consider the general second-order form, from which the previously mentioned filter structures follow directly through suitable parameter choices.

Consider therefore the canonical transfer function
\begin{equation}
	K(z)
	=
	\frac{b_0+b_1z^{-1}+b_2z^{-2}}
	{1+a_1z^{-1}+a_2z^{-2}}.
\end{equation}
Using the denominator of a fixed and stable second-order SISO GOBF (e.g., a second-order Laguerre basis), whose denominator polynomial is denoted by $\phi_{\mathrm{den}}(z),$ a coprime factorization is obtained as
\begin{equation}
	\begin{split}
		U(z)
		&=
		\frac{b_0+b_1z^{-1}+b_2z^{-2}}
		{\phi_{\mathrm{den}}(z)},\\
		V(z)
		&=
		\frac{1+a_1z^{-1}+a_2z^{-2}}
		{\phi_{\mathrm{den}}(z)}.
	\end{split}
\end{equation}

Although stability of the controller $K=UV^{-1}$ is not required by the proposed synthesis framework, it can be enforced by requiring the denominator factor $V$ to be minimum phase. Similarly, minimum phase of $K$ can be imposed by requiring the numerator factor $U$ to be minimum phase. For the second-order parametrization, both properties can be imposed through the Jury stability conditions~\cite{franklin1998digital},
{\small\begin{subequations}\label{eq:jury_conditions}
	\begin{align}
		&1+a_1+a_2>0,\
		1-a_1+a_2>0,\
		1-a_2>0,\\
		&b_0+b_1+b_2>0, \ b_0-b_1+b_2 > 0, \ b_0 - b_2 > 0, \ b_0 > 0.
	\end{align}
\end{subequations}}
Although analogous stability conditions exist for higher-order canonical transfer functions, for example through $D$-stability techniques~\cite{peaucelle2000new}, these quickly become computationally unattractive as the controller order increases. This provides further motivation for the GOBF parametrizations proposed in the previous subsection, whose stability follows directly from construction. Nevertheless, the second-order parametrization is sufficiently rich to encompass many controller structures commonly encountered in practice. Several important examples are summarized in Table~\ref{tab:canonical_filters}.

\begin{table}[t]
	\centering
	\caption{Canonical second-order controller structures and their parametrizations.}
	\label{tab:canonical_filters}
	\renewcommand{\arraystretch}{1.15}
	\begin{tabular}{lll}
		\hline
		Structure & Transfer function\\
		\hline
		PID &
		{\small $K(z) = K_P + K_I\frac{T_s}{1-z^{-1}} + K_D \frac{1-z^{-1}}{T_s (1- \alpha z^{-1})}$}\\
		
		Notch &
		{\small $K(z)= \frac{1-2 r_n \cos (\omega_n)z^{-1} +r_n^2 z^{-2}}{1-2 r_d \cos (\omega_d)z^{-1} +r_d^2 z^{-2}}$} \\
		
		Low-pass &
		{\small $K(z) = \frac{1-a}{1-a z^{-1}}$} \\
		
		Lead &
		{\small $K(z) = \frac{1-\beta z^{-1}}{1-\alpha z^{-1}}$} \\
		
		Lag &
		{\small $K(z) = \frac{1-\alpha z^{-1}}{1-\beta z^{-1}}$} \\
		\hline
	\end{tabular}
\end{table}

The proposed construction extends naturally to MIMO controllers by parametrizing each entry of the matrix-valued coprime factors $V$ and $U$ independently, after which structural constraints such as decentralized or block-structured controllers can be imposed directly on the corresponding coefficients, see Subsection~\ref{subsec:decen}. In practice, however, controllers are often realized as cascaded combinations of elementary filter sections, such as PID, notch, lead, lag, and low-pass filters. Such interconnections generally do not preserve affine dependence of the coprime factors on the individual filter parameters and therefore cannot be incorporated directly into the proposed convex synthesis framework. Developing parametrizations that preserve affinity for such cascaded controller architectures therefore remains an important direction for future research.

Instead, the following subsection considers user-defined controller structures imposed directly on the coprime factors, thereby retaining the affine parametrization required for convex synthesis.

\subsubsection{Non-tunable structure}\noindent\\  \vspace*{-10pt}

\noindent
Lastly, we consider user-defined controller structures of the form
\begin{equation}
	K = K_\mr{o} U V^{-1} K_\mr{i},
\end{equation}
where $K_\mr{o}$ is square and stable, $K_\mr{i}$ is square and minimum phase, and the coprime factors $(U,V)$ remain the optimization variables. The fixed pre- and post-compensators encode user-prescribed structure, while the optimization determines the remaining controller dynamics. This representation can be rewritten as $K=\tilde U\tilde V^{-1}$, where the coprimeness of $(\tilde U,\tilde V)$ now depends on the chosen pre- and post-compensators. Such fixed factors provide a convenient mechanism for embedding application-specific prior knowledge into the controller while preserving the affine dependence of the optimization variables. Typical examples include fixed notch filters, integral action, high-frequency roll-off, resonance inversion in feedforward control, and delay compensation.

Determining whether the resulting factors are coprime is, however, nontrivial and will therefore not be addressed here. Instead, the synthesis approach developed in the following section, provides conditions under which coprimeness is guaranteed as part of the optimization procedure.

Together with the canonical filter structures and direct MIMO parametrizations presented previously, this fixed-factor construction demonstrates the flexibility of the coprime-factor approach. A broad class of practically relevant controller architectures can be represented while preserving the affine parametrization required for convex synthesis. Having established this family of controller parametrizations, we now proceed to develop the proposed iteratively convex synthesis framework.
\section{Convex FRF-Data-Driven Controller Synthesis}\label{sec:syn}
In this section, we develop an IQC-based controller synthesis approach for frequency-response data. The key idea is to reformulate closed-loop stability and performance conditions in a form that is affine in the controller parameters, enabling iterative convex optimization. This is achieved by combining coprime-factor parametrizations with an LFR-based lifting of inverse operators.

%In this section, an IQC-based iteratively convex controller synthesis algorithm is presented. The algorithm, first presented in \cite{bloemers2021frequency}, is adapted for discrete-time systems and sufficiently generalized in terms of controller parametrization and performance notions as outlined in previous sections. In order to derive a complete and tractable synthesis algorithm, we proceed by reviewing (internal) stability notions and conditions of the closed-loop map. Using these notions, we will derive a method for lifting out the inverse of the denominator coprime factorization which is applicable to convex IQC-based stability analysis. More specifically, we will show how feasibility of an FDI implies coprimeness of the controller factors. Lastly, given all the ingredients, we give a full controller synthesis algorithm based only on FRF data of the plant to-be-controller, a chosen controller structure, and a performance metric with associated weighting filters which should give a controller for the desired sensitivity shapes.
\subsection{Closed-loop stability}
Consider again the closed-loop interconnection given in~\eqref{eq:closedloop} and shown in Figure~\ref{fig:FFandFBControlConfig}. Provided that the feedforward controller $F$ is stable, e.g. $F \in \mc{RH}_\infty$, stability of the closed-loop system is a joint property of the feedback controller $K$ and the plant $G$. More specifically, let us define the following stability notion for the interconnection between $K$ and $G$.
\begin{definition}[Internal Stability]
	The feedback controller $K$ is said to internally stabilize the plant $G$ if the system
	\begin{equation}\label{eq:internstab}
		\Matrix{cc}{I & G \\ -K & I},
	\end{equation}
	is invertible in $\mc{RH}_\infty(\mb{E})$, e.g. it has no poles in $\mb{E} \cup \partial \mb{D}$.
\end{definition}
Loosely speaking, internal stability means that the closed-loop is stable and that there are no unstable pole-zero cancellations between $K$ and $G$. In order to analyse internal stability, we define the (output) return difference operator $R = I+GK$ and its characteristic function
\begin{equation}
	\chi(z) = \mr{det}(R(z)) = \frac{\rho_\mr{cl}(z)}{\rho_\mr{ol}(z)},
\end{equation}
where $\rho_\mr{cl}(z)$ denotes the characteristic polynomial of the closed-loop system and $\rho_\mr{ol}(z)$ denotes the characteristic polynomial of the open-loop system. Well-posedness of the closed loop is equivalent to $\chi(e^{j \omega}) \ne 0$ for all $\omega \in \Omega$; this also implies that $R$ is bi-proper. Internal stability is equivalent to well-posedness of the closed-loop and $\mr{wno} \ \mr{det}(\chi(e^{j \omega})) = -p_L$ where $p_L$ is the number of poles of $GK$ in $\mb{E}\cup \partial \mb{D}$~\cite{skogestad2007multivariable}. The winding number, denoted by $\mr{wno}$, of the closed Nyquist contour is evaluated along the closed $\mc{O}$-contour~\cite{pierre1989reformulated} with indentations into the open unit circle $\mb{D}$, as defined in~\cite{desoer1980generalized}, in the counter clockwise direction.

In this paper, we assume a left coprime factorizations of the plant $G = D^{-1} N$ and a right coprime factorization of the controller $K =  U V^{-1}$. A coprime factorization of $G$ can be taken as 
\begin{equation}
(D, N) = ((I+KG)^{-1}, G (I+KG)^{-1}),
\end{equation}
for some $K\in\mc{RH}_\infty^{n_\mr{u} \times n_\mr{y}}(\mb{E})$ that internally stabilizes the closed-loop\footnote{Strictly speaking, this only holds if $G$ is absolutely stabilizable~\cite{vidyasagar2022control} which means that there exists a controller $K \in \mc{RH}_\infty$ that internally stabilizes the closed-loop.}. Next, we define the following transfer matrix
\begin{equation}
	\Lambda = D V + N U,
\label{eq:cf_int_stab}
\end{equation}
which is stable by design and may be interpreted as the solution of the Diophantine equation over $\mc{RH}_\infty(\mb{E})$. The transfer matrix $\Lambda$ captures the interaction between the plant and controller factorizations. In particular, internal stability is equivalent to invertibility of $\Lambda$. The following theorem~\cite{vidyasagar2022control, ball2013interpolation} provides insight into the importance of these transfer matrices.

\begin{theorem}[Internal Stability Criterion]\label{thm:internstab}
	The controller $K$ internally stabilizes the closed loop, if and only if
	\begin{equation}
		\mr{wno} \ \mr{det} (\Lambda)= 0.
	\end{equation}
	where $\Lambda$ is defined as in \eqref{eq:cf_int_stab} and $(D, N)$ denote a left coprime factorization of the plant $G$ and $(V, U)$ denote a right coprime factorization of the controller $K$.
\end{theorem}
This theorem implies that $\Lambda$ is proper and stably invertible in $\mc{RH}_\infty(\mb{E})$. This result is contingent on the fact that $(D, N)$ and $(V, U)$ are coprime factors of the plant and controller, respectively. We now present an example that demonstrates the necessity of coprimeness.
\begin{example}\label{example:cf4stab2}
	Let $G(z) = (z-0.5)^{-1}$ and $K(z) = 1$ for which the following factorizations are used:
	\begin{equation*}
		D = \frac{z-0.5}{z+0.3}, \ N = \frac{1}{z+0.3}, \ V = \frac{z+1.1}{z-0.7}, \ U = \frac{z+1.1}{z-0.7},
	\end{equation*}
	such that $(D, N)$ are coprime but $(V,U)$ are not. The negative feedback interconnection between $G$ and $K$ is internally stable, however, we actually have that
	\begin{equation*}
		\mr{wno} (DV + NU) = 1.
	\end{equation*}
	Hence, we cannot conclude internal stability using the conventional frequency-domain approach, showing the necessity of coprime factorizations.
\end{example}

In the next subsection, we will discuss a convex alternative to circle counting for stability analysis.

\subsection{Convex stability analysis}
In this subsection, we will develop a convex stability analysis result based on $\Lambda$. If $\Lambda$ is invertible in $\mc{RH}_\infty^{\bullet \times \bullet}(\mb{E})$, i.e., minimum-phase, the closed-loop system is internally stable. The following result from~\cite{bloemers2021frequency} gaurantees this.
\begin{lemma}[Convex Invertibility Test]
	Given coprime factorizations of the plant $G = D^{-1}N$ and controller $K =  U V^{-1}$, the function $\Lambda \in \mc{RH}_\infty$, as defined in~\eqref{eq:cf_int_stab}, is invertible in $\mc{RH}_\infty^{\bullet \times \bullet}(\mb{E})$ if and only if there exists a multiplier $\Gamma \in \mc{RH}_\infty$ such that
	\begin{equation}
		\Lambda \Gamma + \Gamma^* \Lambda^* \succ 0 \ \mr{on} \ \partial\mb{D},
	\end{equation}
	is feasible.
\end{lemma}
The above result can be interpreted as a passivity condition, which admits a convex formulation. If the controller parameters for $(V, U)$ are given, these are convex conditions. If the controller parameters are unknown, they can be absorbed into the multiplier such that the FDI becomes
\begin{equation}\label{eq:NonParamVU}
	\mr{He}\left( D \Gamma_V + N \Gamma_U\right) \succ 0 \ \mr{on} \ \partial\mb{D},
\end{equation}
such that $[V \ U] \Gamma = [\Gamma_V \ \Gamma_U]$ and $\mr{He}(X) := X^* + X$. Hence, stability can be analyzed in a convex manner, however this result does not provide a way to realize an implementable controller $K$ from $\Gamma_V$ and $\Gamma_U$, but can be used as a non-parametric initialization. 

In order to use affine controller coprime parametrizations, which leads to realizable controllers while retaining convexity, we will now focus directly on stability of $\Lambda^{-1}$. This function is not an affine function of the controller parameters which is why it is not directly suitable for convex optimization. Following the observation in~\cite{bloemers2021frequency}, we employ a different LFR formulation of the inverse function which we will denote as \emph{inverse lifting}. Consider the stable transfer matrix
\begin{equation}
	M:=\Matrix{cc}{I - \Lambda & C \\ B & A} \in \mc{RH}_\infty,
\end{equation}
where $A, B, C, \Lambda$ are compatible stable transfer matrices with $\Lambda$ invertible. Then
\begin{equation}\label{eq:InvLift}
	A+ B\Lambda^{-1} C = \Delta_I \star M,
\end{equation}
where $\Delta_I = I$. This identity allows us to represent inverse operators as interconnections with the identity, enabling the use of IQC-based analysis which we emphasize by using the notation $\Delta_I = I$. Hence, we will be interested in the interconnection $\Lambda^{-1} = \Delta_I \star (I - \Lambda)$ for robust stability analysis.

By pulling out the uncertainty $\Delta_I$, e.g., the identity matrix $I$, in an LFR formulation, we have made the interconnected system $I - \Lambda$ an affine function of the controller parameters $V$ and $U$. In order to use the IQC framework for the interconnection $\Delta_I \star (I - \Lambda),$ we have to define a suitable multiplier class $\boldsymbol{\Pi}_I$ that precisely specifies the singleton uncertainty set $\boldsymbol{\Delta}_I = I$. This will not be derived in this paper since this has already been established in~\cite{bloemers2021frequency}, so we will simply provide the multiplier class:
\begin{equation}
	\boldsymbol{\Pi}_I := \Bigg\{ \Pi \in \boldsymbol{\Pi}_\mr{in} \ \Bigg|  \ \Matrix{c}{I \\ I}^* \Pi \Matrix{c}{I \\ I} \succeq 0 \ \mr{on} \ \partial \mb{D} \Bigg\}.
\end{equation}
%So, for the stability multiplier $\Pi^\mr{s}\in\boldsymbol{\Pi}_I$, we have to respect the following convex constraints:
%\begin{equation}
%	\Matrix{ccc}{\Pi^\mr{s}_\mr{qq} & 0 & 0 \\ 0 & -\Pi^\mr{s}_\mr{\rho\rho} & 0 \\ 0 & 0 & \Pi^\mr{s}_\mr{qq} + \Pi^\mr{s}_\mr{\rho \rho} + \Pi^\mr{s}_\mr{q \rho} + (\Pi^\mr{s}_\mr{q \rho})^*}\succ 0 \ \mr{on} \ \partial \mb{D}.
%\end{equation}
The following theorem is used to analyze robust stability between $\Delta_I$ and $I - \Lambda$ and provides a link with internal stability.
\begin{theorem}[IQC-based Internal Stability]
	If there exists an IQC-multiplier $\Pi^\mr{s}\in\boldsymbol{\Pi}_I$ that certifies robust stability for the feedback interconnection between $I-\Lambda$ and all $\Delta_I \in \boldsymbol{\Delta}_I$. Then the system $\Lambda^{-1}$ is internally stable.
\end{theorem}
\begin{proof}
	Well-posedness of the interconnection means that $I-\Delta_I(I-\Lambda) = \Lambda$ is causally invertible. The existence of a $\Pi^\mr{s} \in \boldsymbol{\Pi}_I$ certifies that $\Delta_I \star (I-\Lambda)$ is robustly stable for all $\Delta \in \boldsymbol{\Delta}_I$. These two conditions imply that $\Lambda$ is stably invertible in $\mc{RH}_\infty$ which is equivalent to $\mr{wno} \ \mr{det} (\Lambda) = 0$ by Theorem~\ref{thm:internstab}. Assuming that $(D, N)$ is coprime, $\mr{wno} \ \mr{det} (\Lambda) = 0$ also guarantees that $(V, U)$ are coprime and, in turn, that the system $\Lambda^{-1}$ is internally stable.
\end{proof}

Such a multiplier will also be referred to as a stability certificate. In the next subsection, it will be shown how this IQC-based stability result can be extended to also include robust performance as presented in Section~\ref{sec:perf}.

\subsection{Closed-loop reformulation for robust performance analysis}

Previously, we derived a convex robust stability analysis condition that guarantees internal stability. We now incorporate the performance formulation developed in Section~\ref{sec:perf} to obtain a unified robust stability and performance synthesis approach. To this end, the stability multiplier $\Pi^\mr{s} \in \boldsymbol{\Pi}_I$ and the performance multiplier $\Pi^\mr{p} \in \boldsymbol{\Pi}_\mr{p}$ are combined through the aggregated multiplier
\begin{equation}\label{eq:PiAug}
	\Pi
	=
	\operatorname{daug}(\Pi^\mr{s},\Pi^\mr{p})
	:=
	\Matrix{cc:cc}{
		\Pi_\mr{qq}^\mr{s} & 0 &
		\Pi_\mr{qp}^\mr{s} & 0 \\
		0 & Q_\mr{p}^{-1} &
		0 & 0 \\
		\hdashline
		(\Pi_\mr{qp}^\mr{s})^* & 0 &
		\Pi_\mr{pp}^\mr{s} & 0 \\
		0 & 0 &
		0 & -R_\mr{p}},
\end{equation}
certifying both robust stability and performance.

To incorporate the aggregated multiplier into the synthesis approach, we first express the closed-loop in terms of coprime factorizations.  For the sensitivity functions~\eqref{eq:ST} considered in this paper and depicted in Fig.~\ref{fig:FFandFBControlConfig}, the mixed-sensitivity closed-loop admits the coprime-factor representation
\begin{equation}
	T_\mr{zw}
	=
	\Matrix{cc}{S & -SG \\ KS & KSG}
	=
	\Matrix{c}{V \\ U}
	\Lambda^{-1}
	\Matrix{cc}{D & N},
\end{equation}
where $\Lambda$ is defined in~\eqref{eq:cf_int_stab}. Alternatively, the reference-to-error sensitivity
\[
(I+GK)^{-1}(I-GF),
\]
which incorporates feedforward as the coprime factorization $F = Y^{-1} X$, can be expressed as
\begin{equation}
	T_\mr{zw}
	=
	\Matrix{cc}{S & -SG}
	\Matrix{cc}{Y & 0 \\ 0 & Y}^{-1}
	\Matrix{c}{Y \\ X},
\end{equation}
assuming the feedback controller is fixed. Simultaneous feedback--feedforward co-design is obtained through
\begin{equation}
	T_\mr{zw}
	=
	\Matrix{cc}{V & 0}
	\Matrix{cc}{\Lambda & N \\ 0 & Y}^{-1}
	\Matrix{c}{D \\ X}.
\end{equation}
Finally, the generalized plant framework of~\cite{zhou1996robust}, in which the closed-loop is represented as $T_\mr{zw}=P\star K$, can be incorporated in the same manner, but is beyond the scope of this paper. Applying the inverse lifting~\eqref{eq:InvLift} introduced in the previous subsection then directly yields the corresponding inverse-lifted representation $T_\mr{zw} = \Delta_I \star M$ where $M$ is stable and an affine function of controller parameters collected in $\theta$. We are now in a position to present the proposed iteratively convex synthesis approach.

\subsection{Controller synthesis algorithm}

%Convex robust performance analysis is not convex in the controller parameters $\theta$ and the stability multiplier $\Pi^\mr{s}$. However, 
If a fixed controller parameter vector $\theta$ is known that internally stabilizes the closed-loop system, performance analysis of the closed loop for the given controller is convex in the aggregated multiplier $\Pi$ by using a slack variable as shown in Lemma~\ref{lemma:Slack}. Therefore, the following optimization problem can be solved to obtain a stability multplier $\Pi^\mr{s}$:
\begin{equation}\label{eq:multsyn}
	\begin{gathered}
		\underset{S, \ \Pi^\mr{s} \in \boldsymbol{\Pi}_I, \ \Pi^\mr{p} \in \boldsymbol{\Pi}_\mr{p}}{\mathrm{minimize}} \ \ \ J(\Pi^\mr{p}) \ \ \  \mr{subject \ to}\\
		(\bullet)^* \Matrix{cccc}{\Pi_\mr{qq}^\mr{s} & 0 & \Pi_\mr{q p}^\mr{s} & 0 \\ \bullet & S & 0 & 0 \\ \bullet & \bullet & \Pi_\mr{pp}^\mr{s} & 0 \\ \bullet & \bullet & \bullet & -R_\mr{p}} \Matrix{c}{M \\ I} \prec 0, \ \Matrix{ccc}{Q_\mr{p} & I \\ \bullet & S} \succ 0.%\\
		%\Pi_\mr{qq}^\mr{s} \succ 0, \ \Pi^\mr{s}_{\rho \rho} \succ 0 \ (\bullet)^* \Matrix{cc}{\Pi_\mr{qq}^\mr{s} & -\Pi_\mr{q \rho}^\mr{s}\\ \bullet & -\Pi_\mr{\rho \rho}^\mr{s} } \Matrix{c}{I \\ I} \succ 0
	\end{gathered}
\end{equation}
Similarly, given a fixed stability multiplier $\Pi^\mr{s}$ that certifies robust stability of the closedloop, Schur's complement can be used to turn the previous optimization program into a convex problem in the structured controller parametrization $\theta\in\Theta$ and the performance multiplier $\Pi^\mr{p}$. Toward this end, introduce the partitioning of the augmented multiplier~\eqref{eq:PiAug} as
\begin{equation}
	\Pi = \Matrix{cc}{\Pi_{11}& \Pi_{12} \\ \Pi_{21} & \Pi_{22}},
\end{equation}
with $\Pi_{12} = \Pi_{21}^*$. Hence, we obtain the following convex optimization problem:
\begin{equation}\label{eq:contrsyn}
	\begin{gathered}
		\underset{\theta \in \Theta, \ \Pi^\mr{p} \in \boldsymbol{\Pi}^\mr{p}}{\mathrm{minimize}} \ \ \ J(\Pi^\mr{p}) \ \ \  \mr{subject \ to}\\
		\Matrix{cc}{\Pi_{22}+M^*(\theta) \Pi_{12} + \Pi^*_{12} M(\theta) & M^*(\theta) \\ M(\theta) & -\Pi_{11}^{-1}} \prec 0 \ \mr{on} \ \partial \mb{D}.%\\
		%\Pi_\mr{qq}^\mr{s} \succ 0, \ \Pi^\mr{s}_{\rho \rho} \succ 0 \ (\bullet)^* \Matrix{cc}{\Pi_\mr{qq}^\mr{s} & -\Pi_\mr{q \rho}^\mr{s}\\ \bullet & -\Pi_\mr{\rho \rho}^\mr{s} } \Matrix{c}{I \\ I} \succ 0
	\end{gathered}
\end{equation}
Starting from any internally stabilizing controller parametrization $\theta_0 \in \Theta$ we can iteratively search for a stability multiplier $\Pi^\mr{s}$ followed by a new controller parametrization. These steps can be taken until convergence to a local minima or another stopping criterion is met. This synthesis approach is summarized in Algorithm~\ref{alg:syn}.
\algrenewcommand\algorithmicensure{\textbf{Return:}}
\begin{algorithm}[t]
	\caption{Iterative IQC-Based Controller Synthesis}\label{alg:syn}
	\begin{algorithmic}[2]
		\Require Controller parametrization given by $\Theta$, initial stabilizing controller $K_0 = U_0V_0^{-1}$ with $\theta_0 \in \Theta$, plant coprime factors $(D,N)$ satisfying $G=D^{-1}N$, lifted interconnection $M=M(D,N,U,V)$, performance index $p\in\{1,2,\infty\}$
		\State {\bf solve} SDP \eqref{eq:multsyn} under fixed $\theta_{0}$, giving $\Pi_{0}$
		\State {\bf set} $J^* \gets J(\Pi_{\mc{H}_p})$ with $\Pi_{0},\theta_{0}$, $i\leftarrow 1$
		\Repeat
		\State {\bf solve} SDP \eqref{eq:contrsyn} under fixed $\Pi^\mr{s}_{i-1}$, giving $\theta_{i}$
		\State {\bf solve} SDP \eqref{eq:multsyn} under fixed $\theta_{i}$, giving $\Pi_{i}$
		%	\begin{equation*}
			%		\begin{aligned}
				%			&\text{minimize} \quad && J(\Pi_{\mathcal{H}_p}) \quad \text{over } \Pi \\
				%			&\text{subject to} \quad && \text{FDI}~\eqref{eq:MultIQCite}, \quad J(\Pi_{\mathcal{H}_p}) < J^*
				%		\end{aligned}
			%	\end{equation*}
		\State {\bf update}  $J^* \gets J(\Pi_{\mathcal{H}_p})$ with $\Pi_{i},\theta_{i}$
		%		\State {\bf solve} SDP \eqref{eq:ContIQCite} under fixed $\Pi_i$, giving $\theta_{i}$
		%		\State {\bf update}  $J^* \gets J(\Pi_{\mathcal{H}_p})$ under $\Pi_i,\theta_i$
		%		\State Solve the SDP:
		%		\begin{equation*}
			%			\begin{aligned}
				%				&\text{minimize} \quad && J(\Pi_{\mathcal{H}_p}) \quad \text{over } (V, U, \Pi_{\mathcal{H}_p}) \\
				%				&\text{subject to} \quad && \text{FDI}~\eqref{eq:ContIQCite}, \quad J(\Pi_{\mathcal{H}_p}) < J^*
				%			\end{aligned}
			%		\end{equation*}
		\State {\bf update} controller $K \gets U_iV_i^{-1}$ based on $\theta_i$
		\State {\bf update} index $i \gets i+1$
		%		\State Update cost: $J^* \gets J(\Pi_{\mathcal{H}_p})$
		\Until{$J^*$ converged}
		\Ensure Controller $K$, ensuring robust stability and performance specs 
		%		\Ensure Controller $K = UV^{-1}$ satisfies robust stability and performance specifications.
	\end{algorithmic}
\end{algorithm}

\subsection{From Controller Synthesis to Controller Design}

The results presented thus so provide a computational framework for synthesizing controllers directly from FRF data. However, practical deployment on high-performance mechatronic systems requires more than the ability to solve a single synthesis problem.

Although Algorithm~\ref{alg:syn} generates a sequence of convex optimization problems, the underlying controller synthesis problem remains non-convex due to the coprime-factor parametrization and the iterative linearization procedure. Consequently, convergence to a particular solution depends on the chosen initialization, controller parametrization, and performance specifications. Different choices may therefore lead to substantially different local optima.

This observation becomes particularly relevant for high-precision motion systems such as the wirebonder, where performance objectives are often conflicting. Increasing bandwidth, enforcing robustness margins, suppressing cross-coupling, and maintaining implementable controller complexity simultaneously introduce a large design space that cannot be explored reliably through a single synthesis step.

From an industrial perspective, controller design therefore becomes a progressive refinement process rather than a one-shot optimization problem. Engineers typically begin with a simple stabilizing controller, establish achievable performance levels, and subsequently increase controller flexibility and performance requirements.

Motivated by this practical reality, the next section introduces a bootstrapped IQC-based design methodology. The proposed workflow combines the synthesis framework developed in this paper with a structured progression of controller parametrizations and performance specifications, enabling systematic exploration of the achievable performance limits of a given system. This bootstrapped workflow will be demonstrated on an industrial wirebonder.

\section{Bootstrapped IQC-Based Controller Design}\label{sec:design}

In this section, the proposed synthesis approach is applied to the motion stage of an industrial wirebonder. First, the experimental platform is introduced together with how the FRF of the plant was estimated from measured data. Subsequently, the proposed methodology with a bootstrapping process is used to synthesize a sequence of increasingly sophisticated feedback and feedforward controllers. Finally, the resulting controllers are implemented and experimentally validated on the physical setup.

\subsection{Industrial Motion Stage of a Wirebonder}

\begin{figure}[t]
	\centering
	\includegraphics[width=0.25\linewidth]{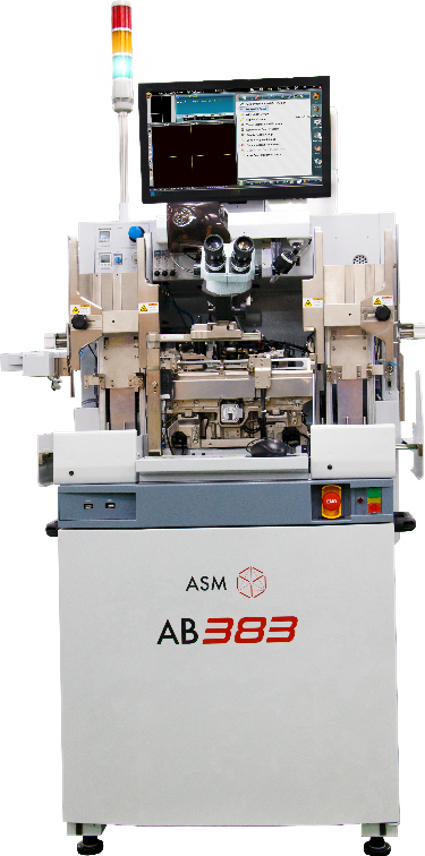}
	\hfill
	\includegraphics[width=0.73\linewidth]{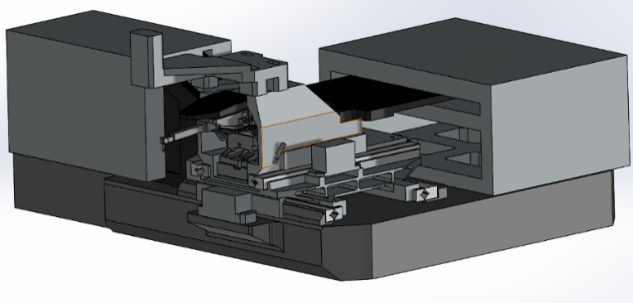}
	\caption{Industrial wirebonder setup (ASM AB383) used for experimental validation shown on the left and a 3D model of the isolated motion stage shown on the right.}
	\label{fig:commercWirebonder}
\end{figure}

The experimental platform considered throughout this paper is the motion stage of a commercial ASM AB383 wirebonder, shown in Fig.~\ref{fig:commercWirebonder}. The motion stage provides the primary positioning system for the bonding operation. It consists of two stacked planar motion axes actuated by linear motors and equipped with high-resolution position encoders, resulting in a two-input two-output motion control system. The inputs to this MIMO system are the motor currents, while the outputs are the corresponding stage positions. Although the stage exhibits approximately decoupled rigid-body behavior at low frequencies, substantial interaction effects and flexible structural dynamics emerge near the achievable closed-loop bandwidth. These characteristics make the system particularly suitable for evaluating advanced multivariable control methodologies that aim to simultaneously achieve high bandwidth, robust stability, and dynamic decoupling. Consequently, it provides a representative system for assessing the practical applicability of the proposed synthesis methodology while avoiding the additional complexity introduced by higher-level machine functionality.

\begin{figure}[b]
	\centering
	\includegraphics[width=0.75\linewidth]{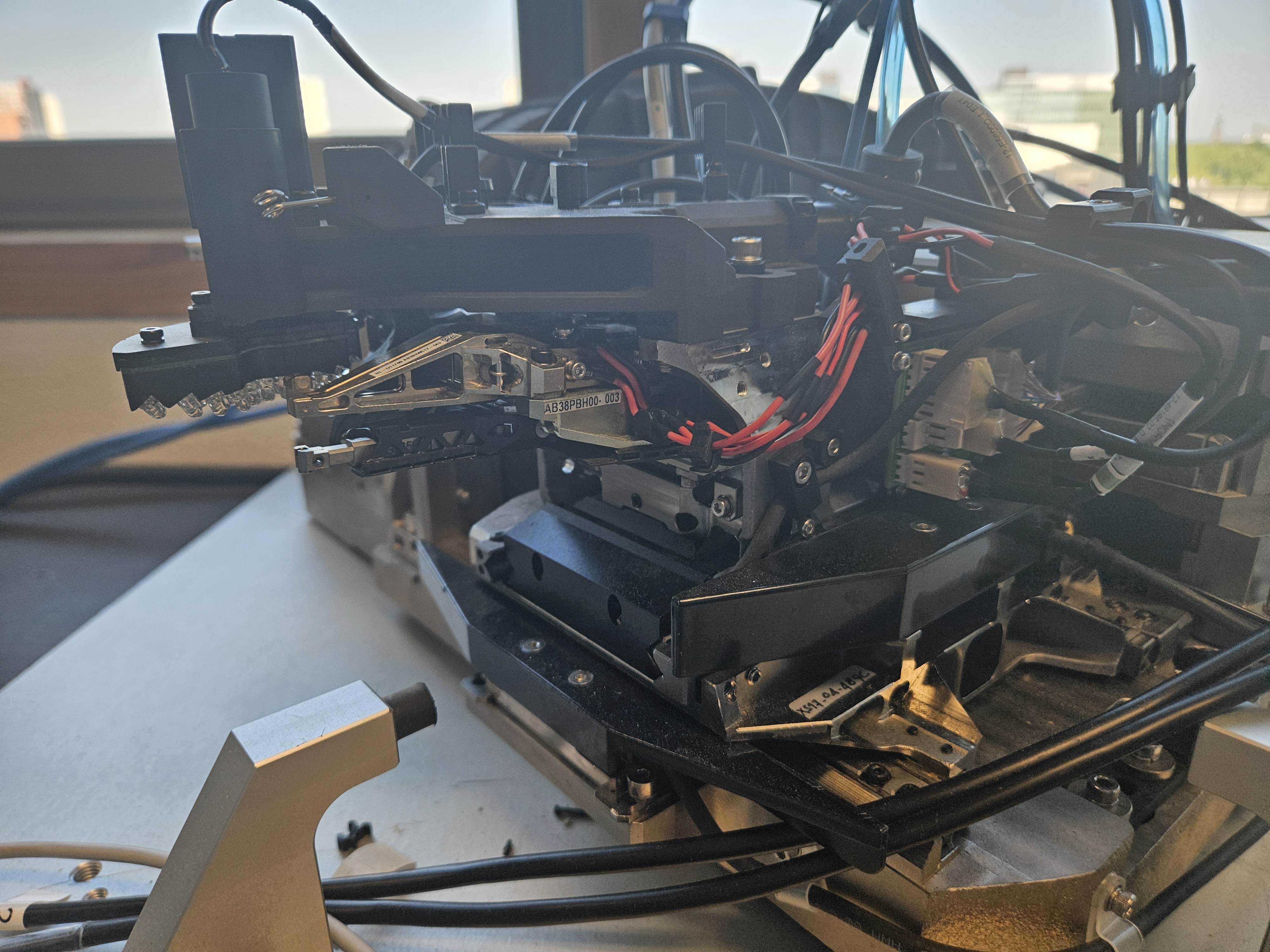}
	\caption{Motion stage of the wirebonder system. The machine is connected to external instrumentation and cooling.}
	\label{fig:wirebonder_lab}
\end{figure}

Controller implementation and experimental validation are performed using a real-time control platform. The controllers are developed in MATLAB/Simulink and deployed through a dSPACE interface, while ControlDesk is used for signal monitoring, parameter adjustment, and data acquisition. The associated laboratory setup is shown in Fig.~\ref{fig:wirebonder_lab}.

%Rather than constructing a parametric model of the system, the subsequent design procedure relies exclusively on experimentally measured frequency response data. The acquisition of these data is discussed in the next subsection.

\subsection{FRF Estimation and Experiment Design}

The proposed synthesis methodology requires a frequency-domain representation of the motion stage. To this end, a nominal FRF estimate of the plant is obtained experimentally using standard frequency-response identification techniques. Since the subsequent synthesis procedure operates directly on the nominal FRF estimate, this non-parametric frequency-domain model is used throughout the remainder of the controller design process.

The wirebonder exhibits position-dependent dynamics due to the (stacked) mechanical configuration of the motion stage. Consequently, different operating points generally correspond to different local frequency response characteristics, which may be interpreted from a frozen-LTI perspective as a family of FRFs, see~\cite{toth2010modeling}. 
%\begin{equation}
%	G_{(x,y)}(e^{j\omega}).
%\end{equation}
Throughout this paper, all identification experiments are performed around the nominal operating point $(x,y)=(0,0)$ and the resulting FRF is denoted by $G(e^{j\omega})$

The FRF measurements are performed in closed-loop using a low-bandwidth diagonal controller. The primary motivation for closed-loop identification is not performance, but maintaining the stage near the selected operating point throughout the experiment. Without feedback, the motion stage may drift away from the desired equilibrium position, thereby altering the local dynamics being identified.

Excitation is applied through orthogonal multisine signals, see~\cite{dobrowiecki2006optimized}, allowing all input--output channels of the MIMO system to be identified simultaneously. The multisine excitation is constructed by exciting $247$ logarithmically spaced frequencies between the fundamental frequency $f_0$ and the Nyquist frequency $f_s/2$. To ensure periodic excitation, these frequencies are rounded to the nearest integer multiple of $f_0$, yielding the frequency grid on which the FRF is experimentally estimated. The plant inputs and outputs are recorded and subsequently processed using the \texttt{RobustLocalPolyAnal.m} routine from the Frequency Domain Toolbox~\cite{pintelon2012system}, implementing the Local Polynomial Method~\cite{gevers2011local} for nonparametric frequency-response estimation. Using this procedure, ten independent multisine realizations were generated. Each realization was repeated ten times and subsequently averaged to improve the quality of the estimated FRF. The exact sequence length and sampling frequency are omitted for confidentiality reasons. Subsequently, the measured FRF is linearly interpolated onto a logarithmically spaced grid of $740$ frequency points spanning the same frequency range, without the integer-frequency constraint. This interpolated representation is used throughout the remainder of the paper for both visualization and controller synthesis.

\begin{figure}[t]
	\centering
	\includegraphics[width=0.95\linewidth]{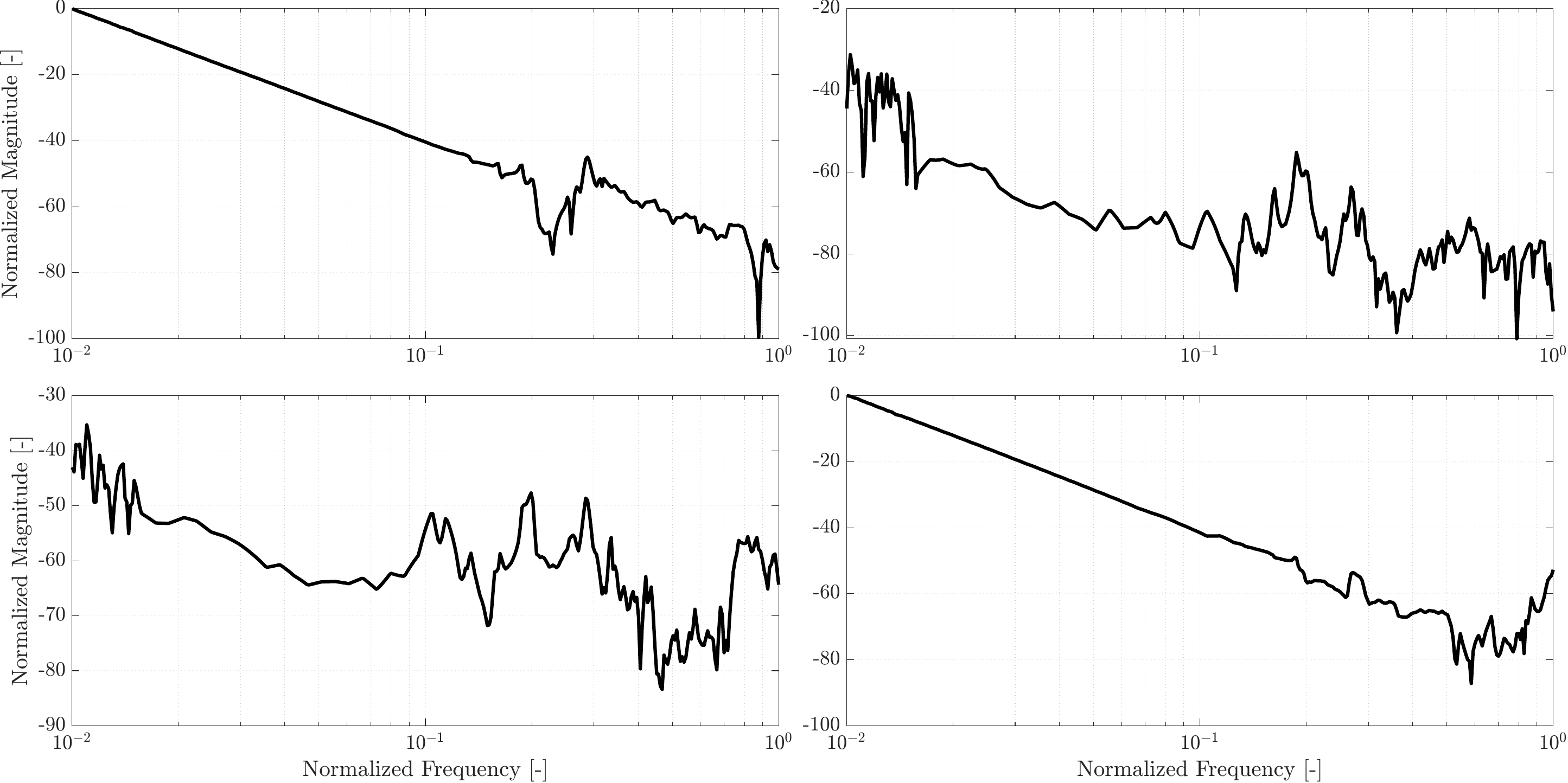}
	\caption{FRF estimation of the isolated motion stage of the wirebonder where the input is current and the output is the stage position. Both the frequency and magnitude axes have been normalized for confidentiality.}
	\label{fig:FRFplot}
\end{figure}

The resulting FRF is shown in Fig.~\ref{fig:FRFplot}. For confidentiality reasons, only normalized magnitude information is presented. Nevertheless, the figure clearly illustrates the dominant diagonal dynamics, the cross-coupling behavior between motion axes, and the flexible dynamics that ultimately limit the achievable closed-loop performance.

\subsection{Bootstrapped IQC-Based Design Methodology}

The synthesis framework developed in the previous sections provides substantial flexibility in both controller parametrization and performance specification. While this flexibility is essential for achieving high closed-loop performance, it also results in a large non-convex design space. Consequently, directly solving the most ambitious synthesis problem is generally not advisable in practice, as the obtained solution may depend strongly on the chosen initialization and may converge to an undesirable local optimum.

This issue is particularly relevant for high-precision mechatronic systems, where multiple competing objectives must be satisfied simultaneously. Increasing bandwidth, enforcing robustness margins, suppressing multivariable interactions, and maintaining implementable controller complexity all introduce additional design freedom and performance trade-offs. As a result, controller design becomes a progressive refinement process rather than a one-shot optimization problem.

To address this challenge, we establish a \emph{bootstrapped design methodology} in which controller flexibility and performance requirements are increased gradually. Starting from a simple stabilizing configuration, each intermediate controller is used to initialize the next synthesis stage. Since all synthesis stages employ compatible coprime-factor parametrizations, the controller obtained in one stage belongs to the admissible controller class of the next stage. As a result, each synthesis problem is initialized from a feasible controller that already satisfies the previously achieved performance level. This approach improves numerical robustness, steers the optimization towards practically meaningful local optima, and provides valuable engineering insight into the mechanisms that limit achievable performance.

\begin{figure}[t]
	\centering
	\includegraphics[width=0.95\linewidth, trim={0pt, 00pt, 00pt, 00pt}, clip]{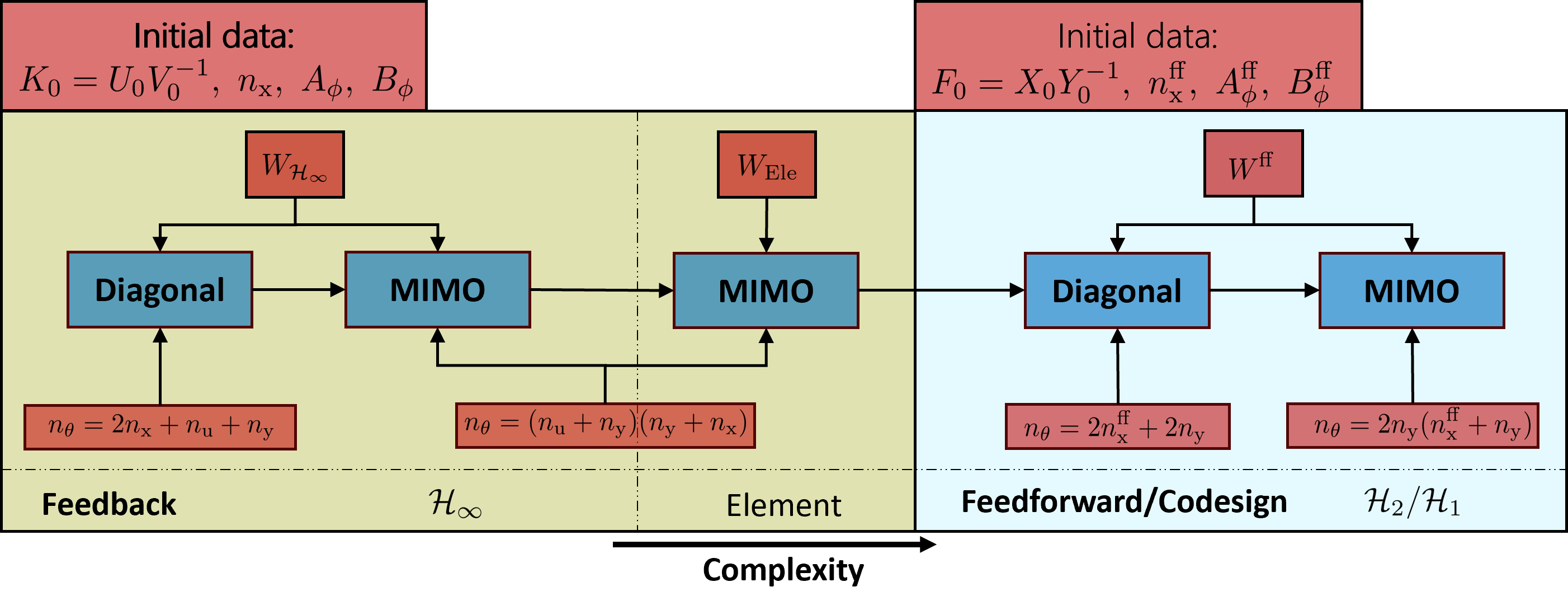}
	\caption{Bootstrapped controller design workflow. Red blocks denote known and fixed data or parameters and the blue blocks denote the iteratively convex synthesis algorithm with bootstrapped initialization. Controller flexibility and performance specifications are progressively increased from diagonal feedback design towards full MIMO feedback-feedforward design.}
	\label{fig:BootstrapSyn}
\end{figure}

An additional advantage of this methodology is that controller complexity is only increased when justified by measurable performance improvements. Consequently, the designer gains a structured understanding of whether observed limitations originate from insufficient controller flexibility, multivariable interactions, robustness requirements, or fundamental plant characteristics such as delay and lightly damped flexible dynamics.

The proposed workflow is summarized in Fig.~\ref{fig:BootstrapSyn} and consists of the following stages.

\subsubsection{Baseline Stabilization}

We begin from any available stabilizing controller. This baseline can be obtained in various ways, such as a non-parametric controller obtained by solving~\eqref{eq:NonParamVU}, an identification-based controller, or an existing industrial design. At this stage, the objective is not performance, but merely to establish a stable closed-loop configuration that serves as a reliable starting point for subsequent refinement.

\subsubsection{Diagonal Performance Design ($\mathcal{H}_\infty$)}

Next, the controller structure is restricted to a diagonal form using~\eqref{eq:diagCon} and an $\mathcal{H}_\infty$ performance objective is specified. Together with the baseline controller, these define the ingredients required to execute Algorithm~\ref{alg:syn}. This step provides a well-conditioned design problem while enabling independent shaping of the dominant loops. As a result, the designer gains insight into the primary dynamics of each channel without the complexity of full multivariable interaction.

\subsubsection{Full MIMO Performance Design ($\mathcal{H}_\infty$)}

The diagonal controller is then used to initialize a full MIMO synthesis step using Algorithm~\ref{alg:syn} in which the structural constraints are relaxed and coupling between channels can be exploited by the optimization. Since the performance specifications remain unchanged, any observed performance improvements originate from the additional controller degrees of freedom rather than modified design objectives.

By starting from a structured solution, the optimization avoids unfavorable local minima while providing insight into the extent to which multivariable interactions limit achievable performance. Moreover, this stage establishes whether the prescribed specifications can be satisfied using a full MIMO controller before introducing more advanced channel-dependent performance objectives.

\subsubsection{Element-Wise Performance Shaping}

After obtaining a satisfactory MIMO design, the performance objective is refined to allow element-wise shaping useing~\eqref{eq:elementPerf}. In contrast to classical approaches based on singular value shaping, this formulation enables direct control over individual input--output channels.

This added flexibility is particularly important in high-precision systems, where performance trade-offs are often channel-specific. For example, it becomes possible to:
\begin{itemize}
	\item selectively suppress specific cross-couplings while preserving diagonal performance,
	\item enforce comparable levels of decoupling across channels,
	\item deliberately relax constraints on certain directions to improve performance elsewhere.
\end{itemize}

In particular, \emph{equal decoupling between channels}---a desirable property in many multivariable motion systems---can be achieved through appropriate choices of element-wise weighting filters. More generally, the framework allows arbitrary channel-dependent trade-offs that cannot be expressed through singular value constraints alone.

This stage effectively provides full control over the distribution of performance across the closed-loop system, enabling targeted mitigation of waterbed effects in a channel-wise manner.

\subsubsection{Optional Feedforward Co-Design}

Finally, feedforward components can be incorporated to further enhance performance, particularly for tracking and disturbance rejection tasks. These elements can be designed jointly with the feedback controller, ensuring consistency with the previously established performance specifications.

\vspace*{5pt}

The proposed workflow can be interpreted as a \emph{progressive refinement strategy}, in which:
\begin{enumerate}
	\item stability is established first,
	\item diagonal performance is shaped,
	\item controller flexibility is subsequently increased through MIMO synthesis,
	\item and fine-grained channel-wise performance shaping is applied last.
\end{enumerate}

This staged approach reflects practical control design practice while leveraging the flexibility of the proposed IQC-based synthesis approach. Importantly, the methodology is \emph{system-independent} and can be applied to a wide range of FRF-based design problems without requiring parametric models.

In the next subsection, the design methodology is applied to the industrial wirebonder introduced previously. The resulting experimental study demonstrates how progressively increasing controller flexibility and performance specification complexity translates into measurable closed-loop performance improvements in practice.

\subsection{Controller Design via Bootstrapped Synthesis}

The bootstrapped design methodology of Fig.~\ref{fig:BootstrapSyn} is now applied to the motion stage of the wirebonder. All controllers are synthesized directly from measured FRF data and satisfy the structural constraints introduced in Section~\ref{sec:controller}, including integral action and high-frequency roll-off.

Table~\ref{tab:designsummary} summarizes the complexity and computational effort associated with each synthesis stage.

\begin{table}[!t]
	\caption{Summary of synthesized controllers.}
	\label{tab:designsummary}
	\centering
	\begin{tabular}{lccccc}
		%\toprule
		Design & Order & $n_\theta$ & Iter. & Time [s] & Performance $\gamma$\\
		%\midrule
		SISO $x$ FB & 12 & 24 & 6 & 100.9 & 1.125\\
		SISO $y$ FB & 12 & 24 & 3 & 50.3 & 1.046\\
		MIMO $\mathcal{H}_\infty$ FB & 24 & 96 & 6 & 175.3 & 0.996\\
		Element-wise FB & 24 & 96 & 10 & 1230.9  & 0.982\\
		Dynamic FF ($x$) & 4 & 8 & 8 & 96.7 & 0.832\\
		Dynamic FF ($y$) & 4 & 8 & 7 & 82.0 & 0.255\\
		%\bottomrule
	\end{tabular}
\end{table}

\subsubsection{Diagonal Feedback Design}

High-precision mechatronic systems are typically designed to exhibit approximately decoupled behavior in the low-frequency range. Although significant coupling may appear near resonances and bandwidth limitations, the dominant rigid-body dynamics can often be treated independently.

Exploiting this property, the first synthesis stage consists of designing two independent SISO controllers, one for each motion axis. These controllers are subsequently implemented as a diagonal multivariable controller. Since the synthesis itself does not impose any multivariable stability criterion, closed-loop stability of the resulting MIMO implementation is verified a posteriori.

\begin{figure}[t]
	\centering
	\includegraphics[width=0.9\linewidth]{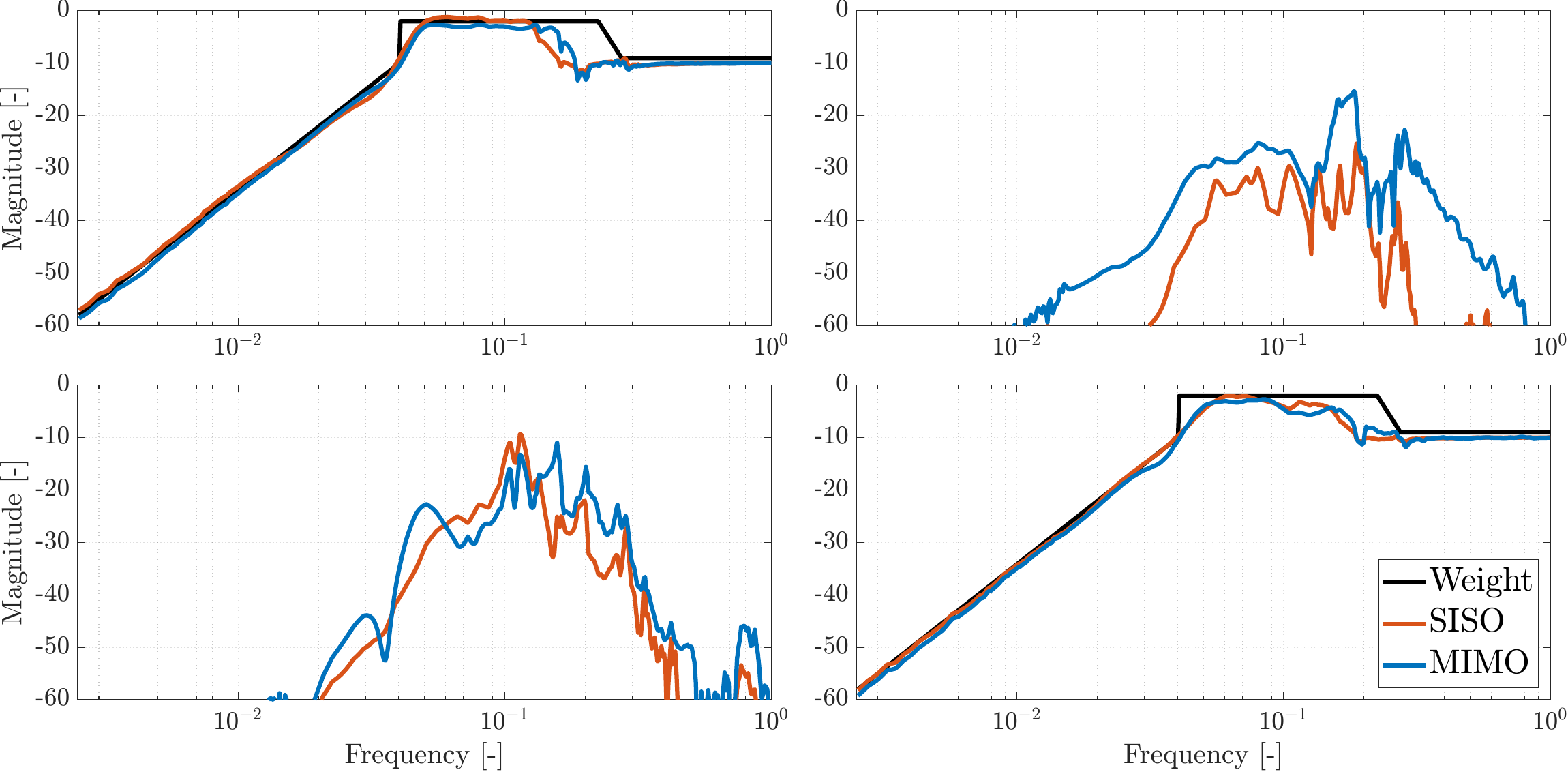}
	\caption{Sensitivity shapes obtained from the designed SISO and MIMO controllers.}
	\label{fig:SISOvsMIMO}
\end{figure}

The synthesis is initialized using the non-parametric controller construction discussed in Section~\ref{sec:syn}. For each axis, a $12^{\mathrm{th}}$-order controller is synthesized using the $\mathcal{H}_\infty$ performance formulation of Section~\ref{sec:perf}. The weighting filters consist of discontinuous magnitude bounds that directly encode the desired sensitivity shape, see Figure~\ref{fig:SISOvsMIMO}. Pole locations are omitted for confidentiality reasons, but are selected near the desired closed-loop bandwidth, which was found to provide a suitable balance between flexibility and numerical conditioning.

The diagonal synthesis stage serves not only as an engineering baseline, but also as a numerically favorable initialization for subsequent synthesis stages. By first optimizing the dominant channels independently, a large portion of the achievable closed-loop performance is already captured before introducing the additional degrees of freedom associated with multivariable coupling. Consequently, the subsequent MIMO synthesis is initialized close to a practically meaningful local optimum while maintaining a well-conditioned optimization problem.

Interestingly, neither SISO controller is capable of achieving $\gamma < 1$. Extensive tuning experiments using alternative initializations and increased controller flexibility produced only marginal improvements. This indicates that the imposed sensitivity specifications are already close to the achievable performance limits of the system. In particular, the dominant limitation is not multivariable coupling, but rather the combination of transport delay and modulus margin requirements, which fundamentally restrict the attainable bandwidth.

Consequently, the diagonal design stage serves two purposes. First, it provides a computationally inexpensive initialization for subsequent synthesis stages. Second, it establishes a practical estimate of the achievable closed-loop performance before multivariable effects are considered.

\begin{figure}[b]
	\centering
	\includegraphics[width=0.9\linewidth]{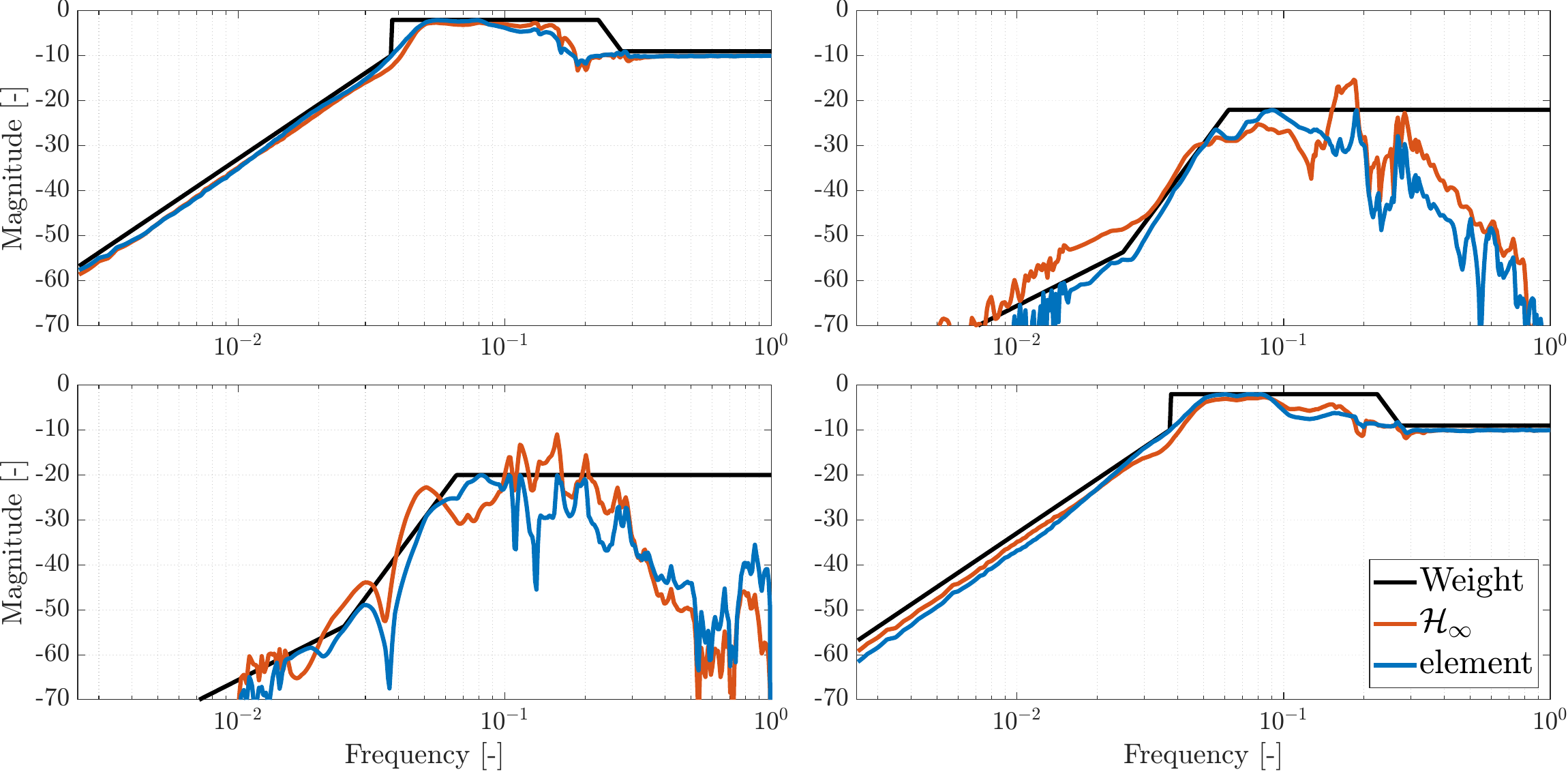}
	\caption{Sensitivity shapes obtained from the designed $\mathcal{H}_\infty$ MIMO and Element-wise MIMO controllers.}
	\label{fig:HinfvsEle}
\end{figure}

\subsubsection{Full MIMO $\mathcal{H}_\infty$ Design}

The diagonal controller is used next as initialization for a full MIMO $\mathcal{H}_\infty$ synthesis. The controller order is increased to $24$, while retaining the same GOBF pole locations, integral-action constraints, high-frequency roll-off constraints, and diagonal weighting filters.

The resulting controller contains $n_\theta=96$ optimization variables, where the additional flexibility originates entirely from the off-diagonal controller entries. Unlike the diagonal design, the full MIMO controller is capable of satisfying the prescribed sensitivity specifications, see Figure~\ref{fig:SISOvsMIMO}.

Figure~\ref{fig:SISOvsMIMO} compares the resulting sensitivity functions. Although the desired singular-value performance is achieved, significant off-diagonal sensitivity terms appear. This behavior is expected, since classical $\mathcal{H}_\infty$ synthesis only shapes the largest singular value of the closed-loop map and therefore does not directly regulate the distribution of performance among individual channels.

The appearance of these coupling terms motivates the transition towards element-wise performance shaping.

\subsubsection{Element-Wise Feedback Design}

The full MIMO $\mathcal{H}_\infty$ controller is subsequently used as initialization for an element-wise synthesis step. The controller parametrization remains unchanged, such that any observed performance differences originate exclusively from the performance specification.

The diagonal sensitivity weights are retained, while two additional off-diagonal weighting filters are introduced to explicitly suppress coupling between motion axes. These filters allow cross-coupling to be penalized without sacrificing the diagonal performance already achieved by the MIMO $\mathcal{H}_\infty$ controller.

Figure~\ref{fig:HinfvsEle} compares the resulting sensitivity functions. Relative to the $\mathcal{H}_\infty$ design, a substantial reduction in off-diagonal sensitivity is achieved while preserving the diagonal sensitivity characteristics. The resulting controller therefore exhibits a significantly more uniform distribution of performance across all channels.

This design stage is computationally more demanding than the previous $\mathcal{H}_\infty$ synthesis, see Table~\ref{tab:designsummary}. Although the individual LMIs are somewhat smaller, approximately four times as many frequency-domain constraints must be enforced, resulting in a substantial increase in computational time.

\subsubsection{Dynamic Feedforward Design}

\begin{figure}[t]
	\centering
	\includegraphics[width=0.9\linewidth]{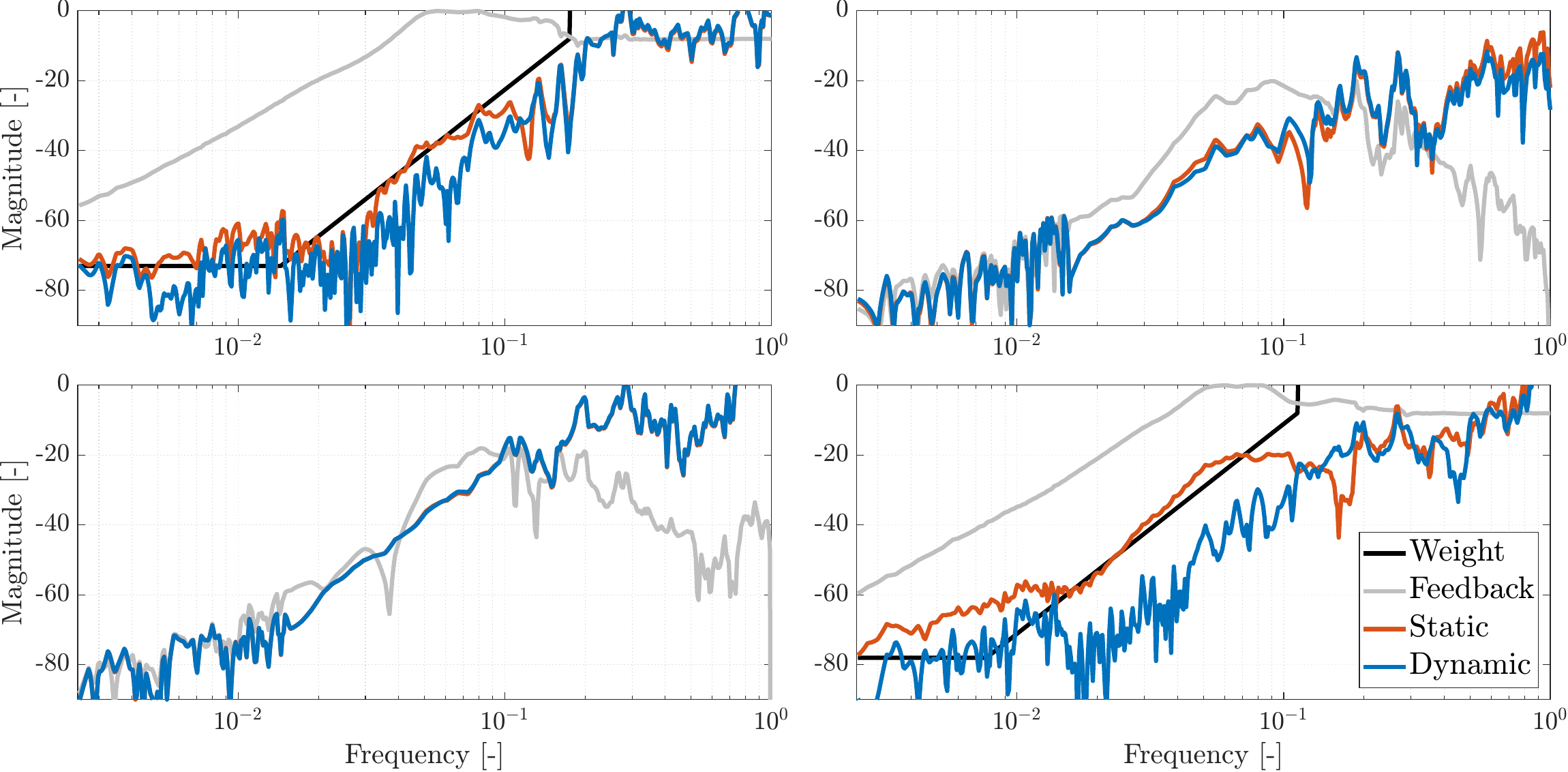}
	\caption{Reference to error sensitivity plots. "Feedback" corresponds to feedback controller design without feedforward, "Static" corresponds to both feedback and static feedforward control, "Dynamic" corresponds to an additional filtered version of the static feedforward.}
	\label{fig:MIMOFF}
\end{figure}

After completing the feedback design, the element-wise controller is fixed and a feedforward controller is synthesized.

A conventional static feedforward strategy of the form
\begin{equation*}
	H = K_a a_t + K_v v_t,
\end{equation*}
is first implemented, where $a_t$ and $v_t$ denote time-advanced acceleration and velocity references, respectively. The parameters $(K_a,K_v,d)$, where $d$ denotes the fractional time advancement, are tuned directly on the experimental setup.

The low-frequency region of the measured FRFs is strongly affected by friction-induced nonlinearities, resulting in poor model quality. Consequently, direct optimization of the static feedforward gains is not considered. Instead, the static feedforward signal is treated as a baseline design and the objective becomes the synthesis of a dynamic correction filter.

The feedforward controller is therefore parametrized as $F = X Y^{-1} H,$ where $H$ denotes the previously tuned static feedforward term. To preserve the low-frequency behavior of the static design, the constraints $X(1)=I$ and $Y(1)=I$ are imposed. Hence, the synthesized filter only modifies the frequency range where performance improvements are expected.

The dynamic feedforward filters are designed independently for both axes using fourth-order coprime factorizations. A weighted $\mathcal{H}_2$ objective is employed instead of an $\mathcal{H}_\infty$ criterion. Since the FRF measurements are affected by noise, average performance across the frequency spectrum is considered more relevant than worst-case behavior at individual frequency points.

Initialization is performed using $(X_0,Y_0)=(I,I)$, such that the initial feedforward controller equals the static feedforward design. Each SISO feedforward filter contains eight optimization variables. Pole locations are selected near frequency regions where performance improvements are considered feasible based on engineering experience with the system.

Figure~\ref{fig:MIMOFF} compares the resulting reference-to-error maps. Successive improvements are observed when moving from feedback-only control to static feedforward and finally to dynamic feedforward compensation.

\subsection{Experimental Validation}

\begin{figure}[!b]
	\centering
	\includegraphics[width=0.9\linewidth]{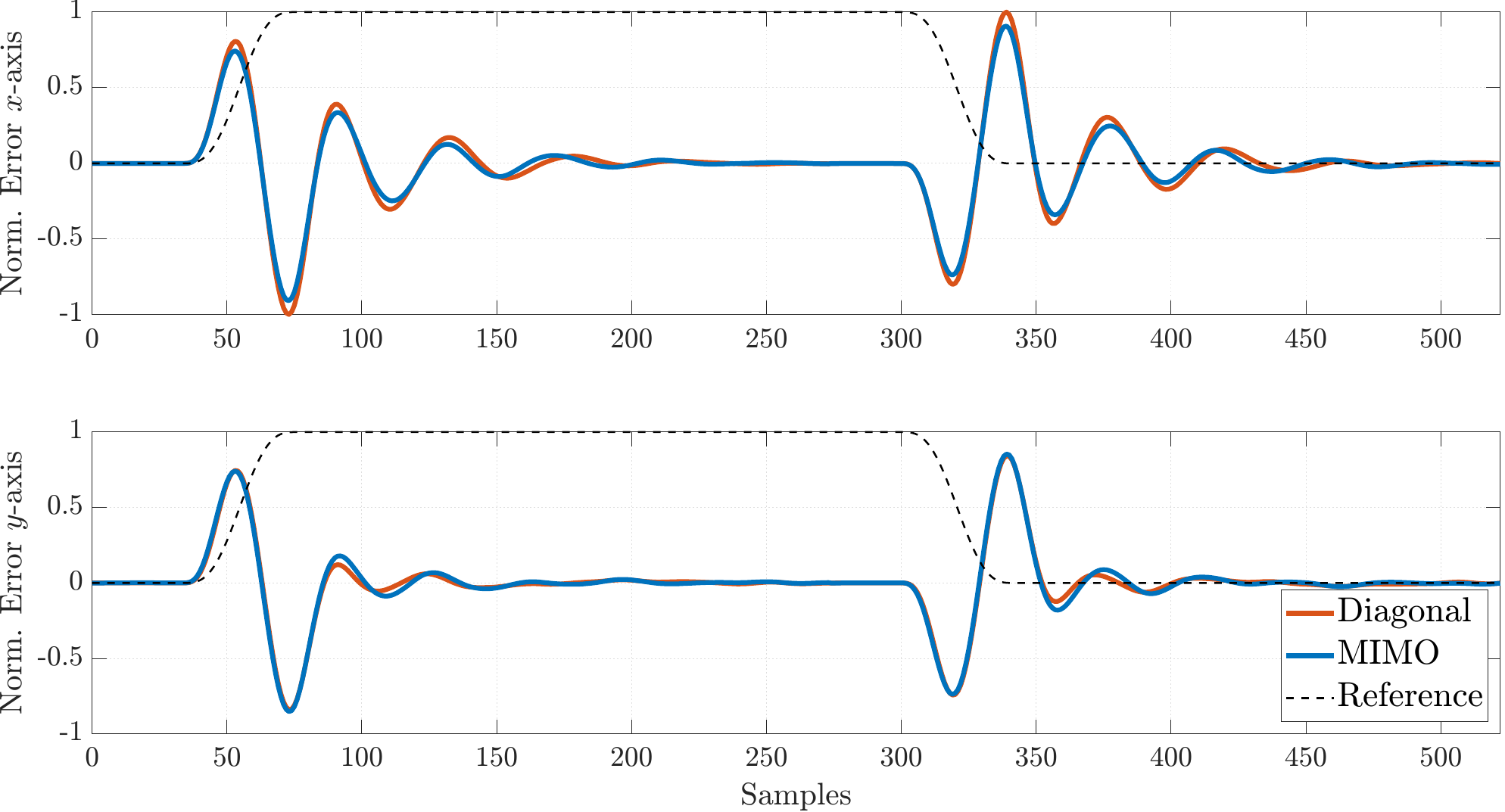}
	\caption{Reference tracking error comparison between diagonal and MIMO $\mathcal{H}_\infty$ controllers for both motion axes.}
	\label{fig:sisovsmimoTD}
\end{figure}

The previous subsection demonstrated how the proposed bootstrapped synthesis methodology progressively increases controller flexibility and performance specifications. The objective of this subsection is to verify that the corresponding improvements predicted by the frequency-domain design procedure are also observed experimentally on the physical wirebonder.

We begin by comparing the diagonal feedback controller against the full MIMO $\mathcal{H}_\infty$ controller. Figure~\ref{fig:sisovsmimoTD} shows the experimentally measured normalized tracking errors obtained for a simultaneous motion of both axes. The top subplot shows the tracking error in the $x$-direction, while the bottom subplot shows the tracking error in the $y$-direction.

The observed responses are largely consistent with the frequency-domain results of Fig.~\ref{fig:SISOvsMIMO}. Although the MIMO controller satisfies the prescribed singular-value specifications and explicitly accounts for multivariable interactions, only limited differences are visible in the time-domain tracking errors. In both motion directions, the transient responses and settling behavior remain comparable. This observation confirms the conclusions drawn from the synthesis results. The dominant limitation on achievable performance is not multivariable coupling, but rather the combination of transport delay and robustness requirements imposed through the modulus margin constraint. Consequently, introducing additional controller degrees of freedom mainly enables satisfaction of the prescribed $\mathcal{H}_\infty$ specifications rather than producing substantial improvements in time-domain performance.

\begin{figure}[!b]
	\centering
	\includegraphics[width=0.9\linewidth]{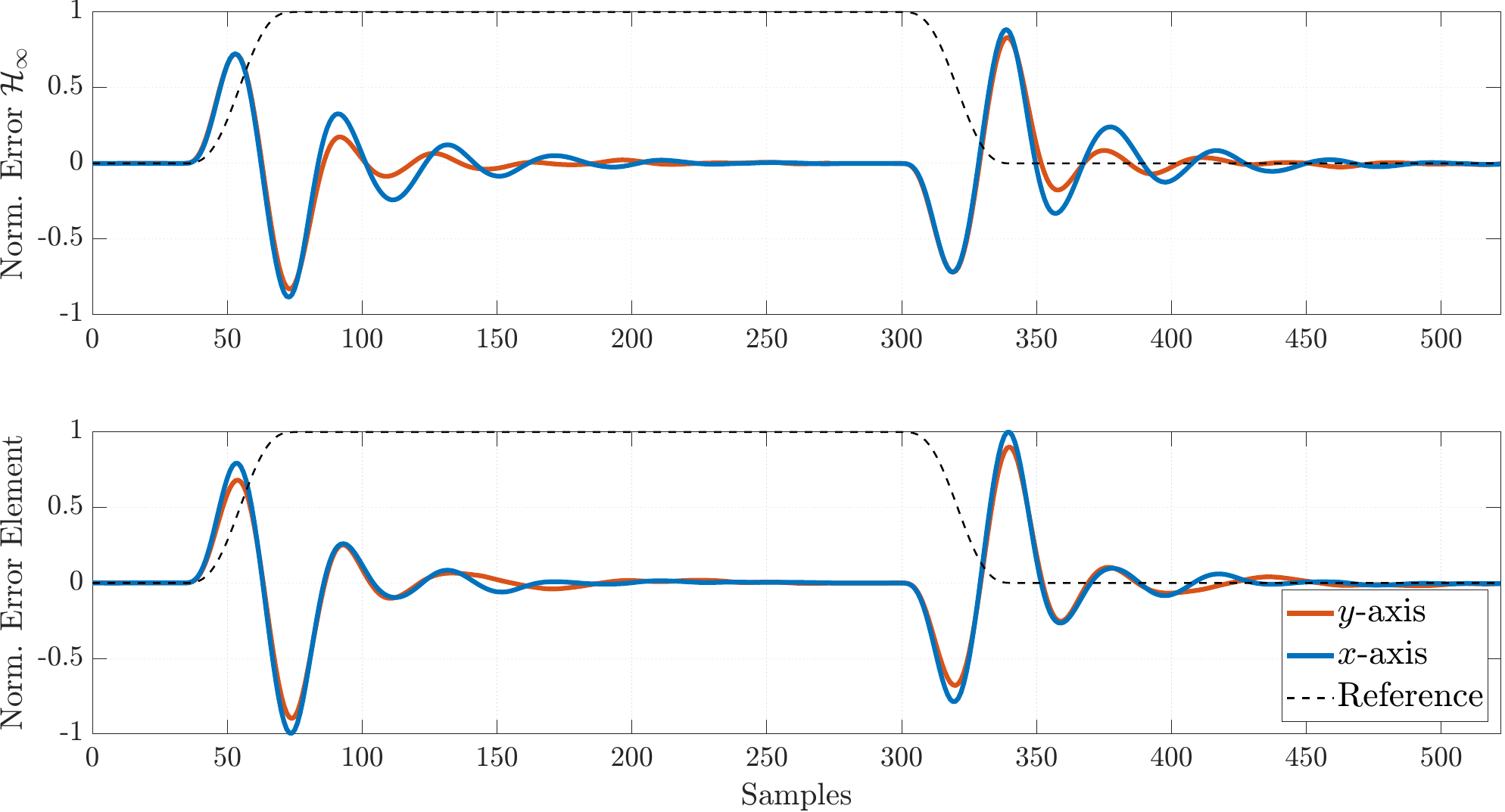}
	\caption{In the top figure the errors of both axes during a motion profile is plotted for the MIMO $\mathcal{H}_\infty$ controller. In the bottom figure the errors of both axes during a motion profile is plotted for the element-wise MIMO controller showing better uniform performance across axes.}
	\label{fig:svdvseleTD}
\end{figure}

This motivates the transition towards element-wise performance shaping. Figure~\ref{fig:svdvseleTD} compares the experimentally measured tracking errors obtained using the MIMO $\mathcal{H}_\infty$ controller and the element-wise controller. In contrast to Figure~\ref{fig:sisovsmimoTD}, where the subplots distinguish between the $x$- and $y$-axes, the subplots in Figure~\ref{fig:svdvseleTD} distinguish between the two controllers. The top subplot shows the responses obtained with the MIMO $\mathcal{H}_\infty$ controller, while the bottom subplot shows the responses obtained with the element-wise controller. In both subplots, the tracking errors of the $x$- and $y$-axes are shown simultaneously.

For the MIMO $\mathcal{H}_\infty$ controller, a clear discrepancy exists between the responses of the two motion axes. One axis settles considerably faster than the other, indicating that the closed-loop performance is distributed unevenly across the system. This behavior is expected since singular-value shaping regulates only the worst-case direction and does not directly constrain the allocation of performance among individual channels.

The element-wise controller exhibits substantially more uniform behavior. Relative to the MIMO $\mathcal{H}_\infty$ controller, the slower axis settles faster while the faster axis settles slightly slower, resulting in nearly identical settling behavior for both axes. Rather than maximizing the performance of a single channel, the element-wise formulation redistributes performance across the system to obtain a more balanced closed-loop response. As a consequence, the overall settling behavior of the multivariable system improves providing higher throughput during bonding. This result directly validates the motivation for introducing element-wise performance shaping and demonstrates that the proposed formulation provides explicit control over the distribution of performance among individual channels.

\begin{figure}[!t]
	\centering
	\includegraphics[width=0.9\linewidth]{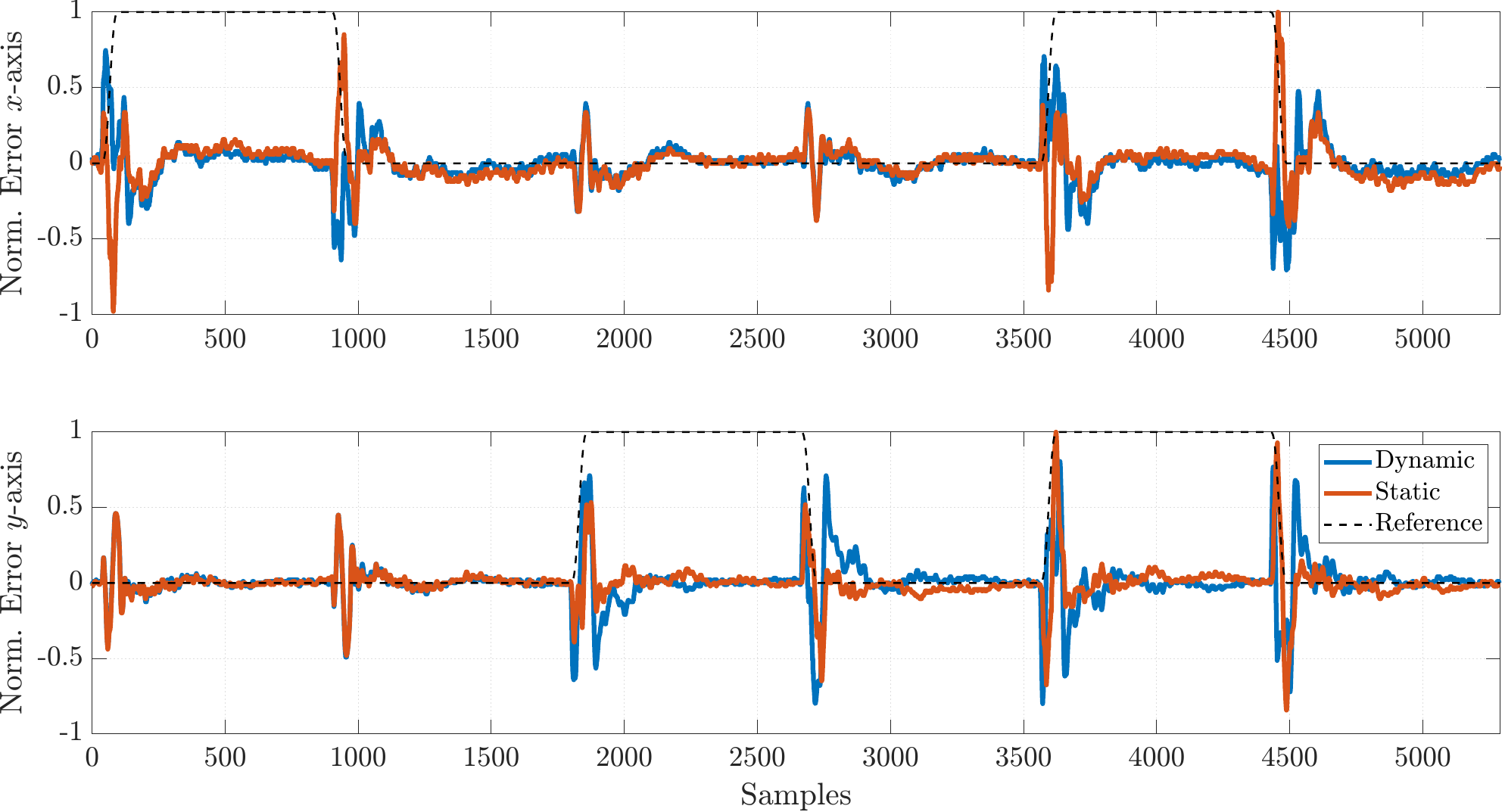}
	\caption{Reference tracking error comparison of the static and dynamic feedforward controllers for three motion profiles: $x$-only, $y$-only, and simultaneous motion. The top subplot shows the $x$-axis tracking error and the bottom subplot shows the $y$-axis tracking error.}
	\label{fig:FFsettling}
\end{figure}

Having established a high-performance feedback controller, the final design stage focuses on feedforward compensation. Figure~\ref{fig:FFsettling} compares the tracking errors obtained using the manually tuned static feedforward strategy and the proposed dynamic feedforward controller. Three representative motion scenarios are considered. First, a motion is performed in the $x$-direction only. Second, a motion is performed in the $y$-direction only. Finally, a simultaneous motion of both axes is executed. The top subplot shows the tracking error in the $x$-direction, while the bottom subplot shows the tracking error in the $y$-direction.

For all considered motion profiles, the dynamic feedforward controller consistently reduces the tracking error relative to the static feedforward strategy. The most pronounced improvement is observed during the settling phase following acceleration and deceleration of the motion stage. These observations are consistent with the frequency-domain results of Fig.~\ref{fig:MIMOFF}, where the dynamic feedforward controller achieved a lower reference-to-error transfer function than the static baseline. The results demonstrate that meaningful performance improvements remain achievable even after the feedback controller has largely reached the practical bandwidth limitations of the system. By augmenting a manually tuned static feedforward design with a low-order dynamic correction filter, additional tracking improvements can be obtained without modifying the carefully designed feedback loop.

Taken together, the experimental results closely mirror the progression of the proposed bootstrapped design methodology. The diagonal controller provides a strong baseline design, the MIMO $\mathcal{H}_\infty$ controller ensures the desired robustness properties, the element-wise controller improves the distribution of performance across channels and achieves more uniform multi-axis behavior, and the dynamic feedforward controller further enhances tracking performance. These results demonstrate that the proposed FRF-based synthesis framework can systematically translate increased controller flexibility into measurable closed-loop performance improvements on an industrial high-precision motion system.

\section{Conclusion}

This paper presented a practical approach for structured MIMO controller design directly from measured frequency-response data. By combining IQC-based convex synthesis with flexible controller parametrizations, the proposed approach accommodates a broad range of feedback and feedforward architectures, performance specifications, and shaping objectives without requiring a parametric plant model forming the main contributions of this paper. The synthesis approach is complemented by a bootstrapped design methodology. By progressively increasing controller flexibility and performance objectives, the resulting workflow provides a practical and systematic route from an initial stabilizing controller to a high-performance multivariable design.

The methodology was validated experimentally on an industrial wirebonder motion system. Starting from diagonal feedback control, successive improvements were obtained through MIMO $\mathcal{H}_\infty$ synthesis, element-wise performance shaping, and dynamic feedforward design. The experimental results demonstrate that the proposed approach translates increased controller flexibility into measurable performance improvements on industrial hardware, making it a practical tool for high-performance mechatronic control design directly from FRF data.
%\section*{Acknowledgements}
%Portions of the manuscript text were prepared with the assistance of ChatGPT (OpenAI) for language refinement and drafting support. All technical content, mathematical derivations, results, and conclusions were developed and verified by the authors.
\bibliographystyle{IEEEtran}
\bibliography{refs}
\newpage

%\section{Biography Section}
%If you have an EPS/PDF photo (graphicx package needed), extra braces are
 %needed around the contents of the optional %argument to biography to prevent
 %the LaTeX parser from getting confused when it sees the complicated
 %$\backslash${\tt{includegraphics}} command within an optional argument. (You can create
 %your own custom macro containing the $\backslash${\tt{includegraphics}} command to make things
 %simpler here.)
 
%\vspace{11pt}

%\bf{If you include a photo:}\vspace{-33pt}
%\begin{IEEEbiography}[{\includegraphics[width=1in,height=1.25in,clip,keepaspectratio]{fig1}}]{Michael Shell}
%Use $\backslash${\tt{begin\{IEEEbiography\}}} and then for the 1st argument use $\backslash${\tt{includegraphics}} to declare and link the author photo.
%Use the author name as the 3rd argument followed by the biography text.
%\end{IEEEbiography}

%\vspace{11pt}

%\bf{If you will not include a photo:}\vspace{-33pt}
%\begin{IEEEbiographynophoto}{John Doe}
%Use $\backslash${\tt{begin\{IEEEbiographynophoto\}}} and the author name as the argument followed by the biography text.
%\end{IEEEbiographynophoto}

\vfill

\end{document}